\documentclass[aps,prb,twocolumn,noshowpacs,showkeys,superscriptaddress,amsmath,amssymb]{revtex4}

\usepackage{graphicx}
\usepackage{dcolumn}
\usepackage{bm}

\begin{document}

\title{Tailoring Native Defects in LiFePO$_{4}$: Insights from First-Principles Calculations}

\author{Khang Hoang}
\affiliation{Center for Computational Materials Science, Naval Research Laboratory, Washington, DC 20375, USA}%
\affiliation{School of Physics, Astronomy, and Computational Sciences, George Mason University, Fairfax, Virginia 22030, USA}%
\author{Michelle Johannes}
\email{michelle.johannes@nrl.navy.mil}
\affiliation{Center for Computational Materials Science, Naval Research Laboratory, Washington, DC 20375, USA}%

\date{\today}

\begin{abstract} 

We report first-principles density-functional theory studies of native point defects and defect complexes in olivine-type LiFePO$_{4}$, a promising candidate for rechargeable Li-ion battery electrodes. The defects are characterized by their formation energies which are calculated within the GGA+$U$ framework. We find that native point defects are charged, and each defect is stable in one charge state only. Removing electrons from the stable defects always generates defect complexes containing small hole polarons. Defect formation energies, hence concentrations, and defect energy landscapes are all sensitive to the choice of atomic chemical potentials which represent experimental conditions. One can, therefore, suppress or enhance certain native defects in LiFePO$_{4}$ via tuning the synthesis conditions. Based on our results, we provide insights on how to obtain samples in experiments with tailored defect concentrations for targeted applications. We also discuss the mechanisms for ionic and electronic conduction in LiFePO$_{4}$ and suggest strategies for enhancing the electrical conductivity.

\end{abstract}
\keywords{lithium iron phosphate, defects, first-principles, polaron, ionic conduction}

\maketitle

\section{\label{sec:intro}Introduction}

Olivine-type LiFePO$_{4}$ is a promising candidate for rechargeable Li-ion battery electrodes.\cite{padhi:1188} The material is known for its structural and chemical stabilities, high intercalation voltage ($\sim$3.5 V relative to lithium metal), high theoretical discharge capacity (170 mAh/g), environmental friendliness, and potentially low costs.\cite{Ellis2010,Manthiram2011} The major drawback of LiFePO$_{4}$ is poor ionic and electronic conduction (with an electrical conductivity of about 10$^{-9}$ S/cm at 298 K)\cite{delacourt:A913} that limits its applicability to devices. While the conduction can be improved by, e.g., making LiFePO$_{4}$ nanoparticles and coating with conductive carbon,\cite{huang:A170,Ellis2007} the high processing cost associated with the manufacturing of carbon-coated LiFePO$_{4}$ nanoparticles may make it less competitive than other materials. Another approach is to dope LiFePO$_{4}$ with aliovalent impurities (Mg, Ti, Zr, Nb) which was reported to have enhanced the conductivity by eight orders of magnitude.\cite{Chung:2002p246} The role of these dopants in the conductivity enhancement, however, remains controversial.\cite{Ravet2003,Wagemaker2008} A better understanding of aliovalent doping, and also better solutions for improving the performance, first requires a deeper understanding of the fundamental properties, especially those associated with native defects, which is currently not available. First-principles density-functional theory (DFT) studies of native point defects and defect complexes in LiFePO$_{4}$ can help address these issues.

It is now generally accepted that LiFePO$_{4}$ is an insulating, large band-gap material in which electronic conduction proceeds via hopping of small hole polarons.\cite{Zhou:2004p101,Maxisch:2006p103,Ellis2006,Zaghib2007} These polarons may be coupled to other defects such as lithium vacancies.\cite{Maxisch:2006p103,Ellis2006} Iron antisites (Fe$_{\rm Li}$) have also been reported to be present in LiFePO$_{4}$ samples.\cite{Yang2002,maier2008,chen2008,Chung2008,Axmann:2009p137,Chung2010} This native defect is believed to be responsible for the loss of electrochemical activity in LiFePO$_{4}$ due to the blockage of lithium channels caused by its low mobility.\cite{Axmann:2009p137,Chung2010} Clearly, native defects have strong effects on the material's performance. Experimental reports on the defects have, however, painted different pictures. Some authors reported evidence of some iron and lithium atoms exchanging sites and forming the antisite pair Fe$_{\rm Li}$-Li$_{\rm Fe}$,\cite{Chung2008,Chung2010} while others determined that Fe$_{\rm Li}$ is formed in association with lithium vacancies ($V_{\rm {Li}}$).\cite{maier2008,Axmann:2009p137} These conflicting reports suggest that the results may be sensitive to the actual synthesis conditions, and indicate that a better understanding of the formation of native defects in LiFePO$_{4}$ is needed in order to produce samples with controlled defect concentrations.

Computational studies of native defects in LiFePO$_{4}$ and related compounds have been reported by several research groups.\cite{Maxisch:2006p103,morgan:A30,Islam:2005p168,Fisher:2008p80,Adams:2010p132,Malik2010} Notably, Maxisch et al.~studied the migration of small hole polarons in LiFePO$_{4}$ using first-principles calculations where the polarons were created both in the absence and in the presence of lithium vacancies.\cite{Maxisch:2006p103} The first systematic study of native defects in LiFePO$_{4}$ was, however, carried out by Islam et al.~using interatomic-potential simulations where they found the antisite pair Fe$_{\rm Li}$-Li$_{\rm Fe}$ to be energetically most favorable.\cite{Islam:2005p168,Fisher:2008p80} Based on results of first-principles calculations, Malik et al.~recently came to a similar conclusion about the antisite pair.\cite{Malik2010} Although these studies have provided valuable information on the native defects in LiFePO$_{4}$, they have three major limitations. First, studies that make use of interatomic potentials may not well describe all the defects in LiFePO$_{4}$. Second, these studies seem to have focused on neutral defect complexes and did not explicitly report the structure and energetics of native point defects as individuals. Third, and most importantly, none of these previous studies have thoroughly investigated the dependence of defect formation energies and hence defect concentrations on the atomic chemical potentials which represent experimental conditions during synthesis.

We herein report our first-principles studies of the structure, energetics, and migration of native point defects and defect complexes in LiFePO$_{4}$. We find that defect formation is sensitive to the synthesis conditions. Native defects can occur in the material with high concentrations and therefore are expected to have important implications for ionic and electronic conduction. We will show how conflicting experimental data on the native defects can be reconciled under our results and provide general guidelines for producing samples with tailored defect concentrations. Comparison with previous theoretical works will be made where appropriate. In the following, we provide technical details of the calculations and present the theoretical approach. Next, we discuss the structural and electronic properties of LiFePO$_{4}$ which form the basis for our discussion of the formation of native defects in the material. We then present results of the first-principles calculations for native point defects and defect complexes, focusing on their formation energies and migration barriers, and discuss the dependence of defect formation energies on the atomic chemical potentials. Based on our results, we discuss the implications of native defects on ionic and electronic conduction, and suggest strategies for enhancing the electrical conductivity. Finally, we end this Article with some important conclusions.

\section{\label{sec;metho}Methodology}

{\bf Computational Details.} Our calculations were based on density-functional theory within the GGA+$U$ framework,\cite{anisimov1991,anisimov1993,liechtenstein1995} which is an extension of the generalized-gradient approximation (GGA),\cite{GGA} and the projector augmented wave method,\cite{PAW1,PAW2} as implemented in the VASP code.\cite{VASP1,VASP2,VASP3} In this work, we used $U$=5.30 eV and $J$=1.00 eV for iron in all the calculations (except otherwise noted), i.e., the effective interaction parameter $U$$-$$J$=4.30 eV (hereafter $U$$-$$J$ will be referred to as $U$ for simplicity). This value of $U$ is the averaged value based on those Zhou et al.~calculated self-consistently for iron in LiFePO$_{4}$ (i.e., Fe$^{2+}$: $U$=3.71 eV) and in FePO$_{4}$ (i.e., Fe$^{3+}$: $U$=5.90 eV), which has been shown to correctly reproduce the experimental intercalation potential of LiFePO$_{4}$.\cite{Zhou:2004p104} It is known that the results obtained within GGA+$U$ depend on the value of $U$. However, we have checked the $U$ dependence in our calculations and find that the physics of what we are presenting is insensitive to the $U$ value for 3.71 eV $\le U \le$ 5.90 eV.  

Calculations for bulk olivine-type LiFePO$_{4}$ (orthorhombic $Pnma$; 28 atoms/unit cell) were performed using a 4$\times$7$\times$9 Monkhorst-Pack $\mathbf{k}$-point mesh.\cite{monkhorst-pack} For defect calculations, we used a (1$\times$2$\times$2) supercell, which corresponds to 112 atoms/cell, and a 2$\times$2$\times$2 $\mathbf{k}$-point mesh. The plane-wave basis-set cutoff was set to 400 eV. Convergence with respect to self-consistent iterations was assumed when the total energy difference between cycles was less than 10$^{-4}$ eV and the residual forces were less than 0.01 eV/{\AA}. In the defect calculations, the lattice parameters were fixed to the calculated bulk values, but all the internal coordinates were fully relaxed. The migration of selected defects in LiFePO$_{4}$ was studied using the climbing-image nudged elastic-band method (NEB).\cite{ci-neb} All calculations were performed with spin polarization and, unless otherwise noted, the antiferromagnetic spin configuration of LiFePO$_{4}$ was used.\cite{rousse2003}

{\bf Defect Formation Energies.} Throughout this Article, we employ defect formation energies to characterize different native point defects and defect complexes in LiFePO$_{4}$. The formation energy of a defect is a crucial factor in determining its concentration. In thermal equilibrium, the concentration of the defect X at temperature $T$ can be obtained via the relation\cite{walle:3851,janotti2009} 
\begin{equation}\label{eq;concen} 
c(\mathrm{X})=N_{\mathrm{sites}}N_{\mathrm{config}}\mathrm{exp}[-E^{f}(\mathrm{X})/k_BT], 
\end{equation} 
where $N_{\mathrm{sites}}$ is the number of high-symmetry sites in the lattice per unit volume on which the defect can be incorporated, and $N_{\mathrm{config}}$ is the number of equivalent configurations (per site). Note that the energy in Eq.~(\ref{eq;concen}) is, in principle, a free energy; however, the entropy and volume terms are often neglected because they are negligible at relevant experimental conditions.\cite{janotti2009} It emerges from Eq.~(\ref{eq;concen}) that defects with low formation energies will easily form and occur in high concentrations.

The formation energy of a defect X in charge state $q$ is defined as\cite{walle:3851} 
\begin{eqnarray}\label{eq;eform}
\nonumber
E^{f}({\mathrm{X}}^q)=E_{\mathrm{tot}}({\mathrm{X}}^q)-E_{\mathrm{tot}}({\mathrm{bulk}})-\sum_{i}{n_i\mu_i}\\
+q(E_{\mathrm{v}}+\Delta V+\epsilon_{F}), 
\end{eqnarray} 
where $E_{\mathrm{tot}}(\mathrm{X}^{q})$ and $E_{\mathrm{tot}}(\mathrm{bulk})$ are, respectively, the total energies of a supercell containing the defect X and of a supercell of the perfect bulk material; $\mu_{i}$ is the atomic chemical potential of species $i$ (and is referenced to the standard state), and $n_{i}$ denotes the number of atoms of species $i$ that have been added ($n_{i}$$>$0) or removed ($n_{i}$$<$0) to form the defect. $\epsilon_{F}$ is the electron chemical potential, i.e., the Fermi level, referenced to the valence-band maximum in the bulk ($E_{\mathrm{v}}$). $\Delta V$ is the ``potential alignment'' term, i.e., the shift in the band positions due to the presence of the charged defect and the neutralizing background, obtained by aligning the average electrostatic potential in regions far away from the defect to the bulk value.\cite{walle:3851} Note that we denote defect X in charge state $q$ as X$^{q}$. For example, Fe$_{\rm Li}^{+}$ indicates that defect Fe$_{\rm Li}$ occurs with charge $q$=+1, which is equivalent to Fe$_{\rm Li}^{\bullet}$ in the Kr\"{o}ger-Vink notation. For a brief discussion on the use of notations, see, e.g., Ref.\cite{vdW2010}.

{\bf Chemical Potentials.} The atomic chemical potentials $\mu_{i}$ are variables and can be chosen to represent experimental conditions. $\mu_{i}$ can, in principle, be related to temperatures and pressures via standard thermodynamic expressions. The chemical potential for O$_{2}$ in oxygen gas, for example, is given by\cite{ong2008} 
\begin{equation}\label{eq;oxygen} 
\mu_{\mathrm{O}_{2}}(T,p)=\mu_{\mathrm{O}_{2}}(T,p_{\circ}) + kT {\rm ln}\frac{p}{p_{\circ}}, 
\end{equation} 
where $p$ and $p_{\circ}$ are, respectively, the partial pressure and reference partial pressure of oxygen; $k$ is Boltzmann's constant. This expression allows us to calculate $\mu_{\mathrm{O}_{2}}(T,p)$ if we know the temperature dependence of $\mu_{\mathrm{O}_{2}}(T,p_{\circ})$ at a particular pressure $p_{\circ}$. In this work, we choose the reference state of $\mu_{\mathrm{O}_{2}}(T,p)$ to be the total energy of an isolated O$_{2}$ molecule ($E_{{\rm O}_{2}}^{\rm tot}$).\cite{Note-1}

The value of $\mu_{i}$ is subject to various thermodynamic limits. For LiFePO$_{4}$, the stability condition requires that 
\begin{equation}\label{eq;stability} 
\mu_{\rm Li}+\mu_{\rm Fe}+\mu_{\rm P}+2\mu_{{\rm O}_{2}}=\Delta H^{f}({\rm LiFePO}_{4}), 
\end{equation} 
where $\Delta H^{f}$ is the formation enthalpy. This condition places a lower bound on the value of $\mu_{i}$. Additionally, one needs to avoid precipitating bulk Li, Fe, and P, or forming O$_{2}$ gas. This sets an upper bound on the chemical potentials: $\mu_{i}$$\leq$0.\cite{walle:3851} There are, however, further constraints imposed by other competing Li-Fe-P-O$_{2}$ phases which usually place stronger bounds on $\mu_{i}$. For example, in order to avoid the formation of Li$_{3}$PO$_{4}$, 
\begin{equation}\label{eq;li3po4} 
3\mu_{\rm Li}+\mu_{\rm P}+2\mu_{{\rm O}_{2}}\leq \Delta H^{f}({\rm Li}_{3}{\rm PO}_{4}). 
\end{equation}

After taking into account the constraints imposed by all possible competing phases, one can define the chemical potential range of Li, Fe, and O$_{2}$ that stabilizes LiFePO$_{4}$ which is, in fact, bound in a polyhedron in the three-dimensional ($\mu_{\rm Li}$, $\mu_{\rm Fe}$, $\mu_{{\rm O}_{2}}$) space. For a given point in the polyhedron, one can determine the remaining variable $\mu_{\rm P}$ via Eq.~(\ref{eq;stability}). In this work, the formation enthalpies of all different Li-Fe-P-O$_{2}$ phases are taken from Ong et al.\cite{ong2008} who have computed the energies using a methodology similar to ours. For example, the calculated formation enthalpy of LiFePO$_{4}$ at $T$=0 K (with respect to its constituents) is $-$18.853 eV per formula unit (f.u.),\cite{ong2008} almost identical to that (of $-$18.882 eV/f.u.) obtained in our calculations. Ong et al.~have also calculated the phase diagrams of the quaternary Li-Fe-P-O$_{2}$ system at 0 K that involve all possible phases between Li, Fe, P, and O$_{2}$. These phase diagrams show LiFePO$_{4}$ is stable over a range of the oxygen chemical potential values, from $-$11.52 eV (where the first Fe$^{2+}$-containing phase appears) to $-$8.25 eV (the last of the Fe$^{2+}$-containing phosphates being reduced).\cite{ong2008} This corresponds to $\mu_{{\rm O}_{2}}$ ranging from $-$3.03 to $-$8.25 eV with respect to our chosen reference ($E_{{\rm O}_{2}}^{\rm 
tot}$).

\begin{figure}
\begin{center}
\includegraphics[width=3.0in]{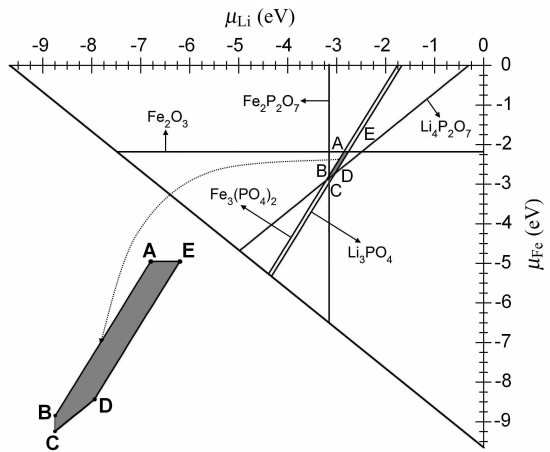}
\end{center}
\vspace{-0.15in}
\caption{Chemical-potential diagram for LiFePO$_{4}$ at $\mu_{{\rm O}_{2}}$=$-$4.59 eV. The $\mu_{{\rm O}_{2}}$ axis extends out of the page. Only phases that can be in equilibrium with LiFePO$_{4}$ are included and the lines delineating these phases define the stability region of LiFePO$_{4}$, here shown as a shaded polygon.}\label{fig;chempot}
\end{figure}

Figure \ref{fig;chempot} shows the slice of the ($\mu_{\rm Li}$, $\mu_{\rm Fe}$, $\mu_{{\rm O}_{2}}$) polyhedron in the $\mu_{{\rm O}_{2}}$=$-$4.59 eV plane, constructed with the calculated formation enthalpies (taken from Ref.\cite{ong2008}) for different Li-Fe-P-O$_{2}$ phases. The shaded area (marked by Points A, B, C, D, and E) shows the range of $\mu_{\rm Li}$ and $\mu_{\rm Fe}$ values where LiFePO$_4$ is stable. Point A, for example, corresponds to equilibrium of LiFePO$_{4}$ with Fe$_{2}$O$_{3}$ and Fe$_{3}$(PO$_{4}$)$_{2}$. At this point in the chemical-potential diagram, the system is close to forming Fe-containing secondary phases (i.e., Fe$_{2}$O$_{3}$ and Fe$_{3}$(PO$_{4}$)$_{2}$) and far from forming Li-containing secondary phases. This can be considered as representing a ``Li-deficient'' environment. Similarly, Point D can be considered as representing a ``Li-excess'' environment, where the system is close to forming Li-containing secondary phases (i.e., Li$_{4}$P$_{2}$O$_{7}$ and Li$_{3}$PO$_{4}$). Note that ``Li-deficient'' and ``Li-excess'' environments in this sense do not necessarily mean that $\mu_{\rm Li}$ in the latter is higher than in the former, as seen in Fig.~\ref{fig;chempot}. Reasonable choices of the atomic chemical potentials should be those that ensure the stability of the host compound. In the next sections we will present our calculated formation energies for various native defects in LiFePO$_{4}$ and discuss how these defects are formed under different experimental conditions.

{\bf Defect Complexes.} Native point defects in LiFePO$_{4}$ may not stay isolated but could instead agglomerate and form defect complexes. For a complex XY consisting of X and Y, its binding energy $E_{b}$ can be calculated using the formation energy of the complex and those of its constituents\cite{walle:3851} 
\begin{equation}
\label{eq;eb} E_{b} = E^{f}({\rm X}) + E^{f}({\rm Y}) - E^{f}({\rm XY}), 
\end{equation} 
where the relation is defined such that a positive binding energy corresponds to a stable, bound defect complex. Having a positive binding energy, however, does not mean that the complex will readily form. For example, under thermal equilibrium, the binding energy $E_{b}$ needs to be greater than the larger of $E^{f}(\rm X)$ and $E^{f}(\rm Y)$ in order for the complex to have higher concentration than its constituents.\cite{walle:3851} For further discussions on the formation of defect complexes, see, e.g., Ref.\cite{walle:3851}.

\section{\label{sec;bulk}Bulk Properties}

Before presenting our results for native defects in LiFePO$_{4}$, let us discuss some basic properties of the pristine compound. Olivine-type LiFePO$_{4}$ was reported to crystallize in the orthorhombic space group $Pnma$ with $a$=10.3377(5), $b$=6.0112(2), and $c$=4.6950(2) {\AA}.\cite{rousse2003} The compound can be regarded as an ordered arrangement of Li$^{+}$, Fe$^{2+}$, and (PO$_{4}$)$^{3-}$ units. Li$^{+}$ forms Li channels along the $b$-axis whereas Fe$^{2+}$ stays at the center of a slightly distorted FeO$_{6}$ octahedron (interwoven with PO$_{4}$ tetrahedra). This simple bonding picture will be very useful when interpreting the structure and energetics of native defects in LiFePO$_{4}$. The calculated lattice parameters are $a$=10.461, $b$=6.061, and $c$=4.752 {\AA}, in satisfactory agreement with the experimental values. The calculated values are slightly larger than the experimental ones as expected since it is well known that GGA tends to overestimate the lattice parameters. The calculated magnetic 
moment for iron (Fe$^{2+}$) is 3.76 $\mu_{\rm B}$, comparable to the experimental value of 4.19(5) $\mu_{\rm B}$ at 2 K.\cite{rousse2003}

\begin{figure}
\begin{center}
\includegraphics[width=3.0in]{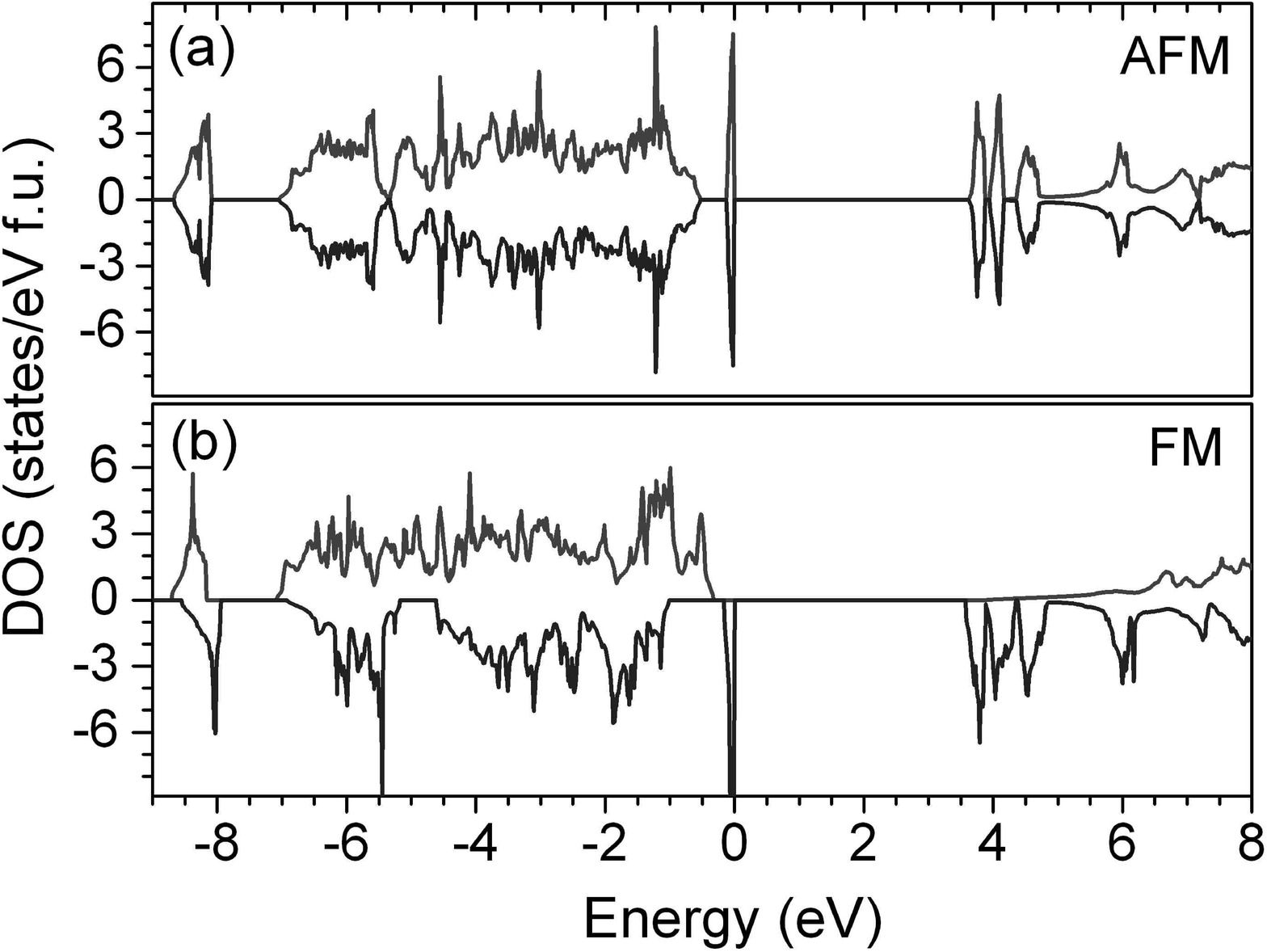}
\end{center}
\vspace{-0.15in}
\caption{Electronic density of states (DOS) of LiFePO$_{4}$ in (a) antiferromagnetic (AFM) and (b) ferromagnetic (FM) spin configurations. The zero of the energy is set to the highest occupied state.}\label{fig;dos}
\end{figure}

Figure \ref{fig;dos} shows the total electronic density of states of LiFePO$_{4}$ in antiferromagnetic (AFM) and ferromagnetic (FM) spin configurations. An analysis of the wavefunctions shows that, in both configurations, the valence-band maximum (VBM) and conduction-band minimum (CBM) are Fe 3$d$ states. Between the highly localized Fe $d$ states just below the Fermi level (at 0 eV) and the lower valence band (which consists predominantly of O 2$p$ and Fe 3$d$ states) there is an energy gap of about 0.40 eV (AFM). The Li 2$s$ state is high up in the conduction band, suggesting that Li donates its electron to the lattice and becomes Li$^{+}$. There is strong mixing between P 3$p$ and O 2$p$ states, indicating covalent bonding within the (PO$_{4}$)$^{3-}$ unit. The calculated band gap is 3.62 and 3.58 eV for AFM and FM spin configurations, respectively, in agreement with previously reported value (of 3.7 eV).\cite{Zhou:2004p101} Experimentally, LiFePO$_{4}$ has been reported to have a band gap of about 3.8$-$4.0 eV, obtained from diffuse reflectance measurements.\cite{Zhou:2004p101,Zaghib2007} The compound is therefore an insulating, large band-gap material.

In the GGA+$U$ framework, the electronic structure can depend on the $U$ value. Indeed, we find that the calculated band gap of LiFePO$_{4}$ is 3.20 and 4.00 eV in the AFM spin configuration for $U$=3.71 and 5.90 eV, respectively, compared to 3.62 eV obtained in calculations using $U$=4.30 eV mentioned earlier. The energy gap between the highest valence band (Fe 3$d$ states) and the lower valence band (predominantly O 2$p$ and Fe 3$d$ states) is also larger for smaller $U$ value: 0.58 and 0.20 eV for $U$=3.71 and 5.90 eV, respectively. However, our GGA+$U$ calculations show that the electronic structure near the band gap region is not sensitive to the choice of $U$ value, for $U$ lying within the range from 3.71 to 5.90 eV. As we illustrate in the next section, knowing the structural and electronic properties, especially the nature of the electronic states near the VBM and CBM, is essential in understanding the formation of native defects in LiFePO$_{4}$.

\section{\label{sec;formation}Formation of Native Defects}

In insulating, large band-gap materials such as LiFePO$_{4}$, native point defects are expected to exist in charged states other than neutral, and charge neutrality requires that defects with opposite charge states coexist in equal concentrations.\cite{peles2007,hoang2009,wilson-short} We therefore investigated various native defects in LiFePO$_{4}$ in all possible charge states. These defects include hole polarons (hereafter denoted as $p^{+}$), lithium vacancies ($V_{\rm Li}$) and interstitials (Li$_{i}$), iron antisites (Fe$_{\rm Li}$), lithium antisites (Li$_{\rm Fe}$), iron vacancies ($V_{\rm Fe}$), and PO$_{4}$ vacancies ($V_{{\rm PO}_{4}}$). We also considered defect complexes that consist of certain point defects such as Fe$_{\rm Li}$-$V_{\rm Li}$ (a complex of Fe$_{\rm Li}$ and $V_{\rm Li}$), Fe$_{\rm Li}$-Li$_{\rm Fe}$ (a complex of Fe$_{\rm Li}$ and Li$_{\rm Fe}$), and 2Fe$_{\rm Li}$-$V_{\rm Fe}$ (a complex of two Fe$_{\rm Li}$ and one $V_{\rm Fe}$).

\begin{figure}
\begin{center}
\includegraphics[width=3.0in]{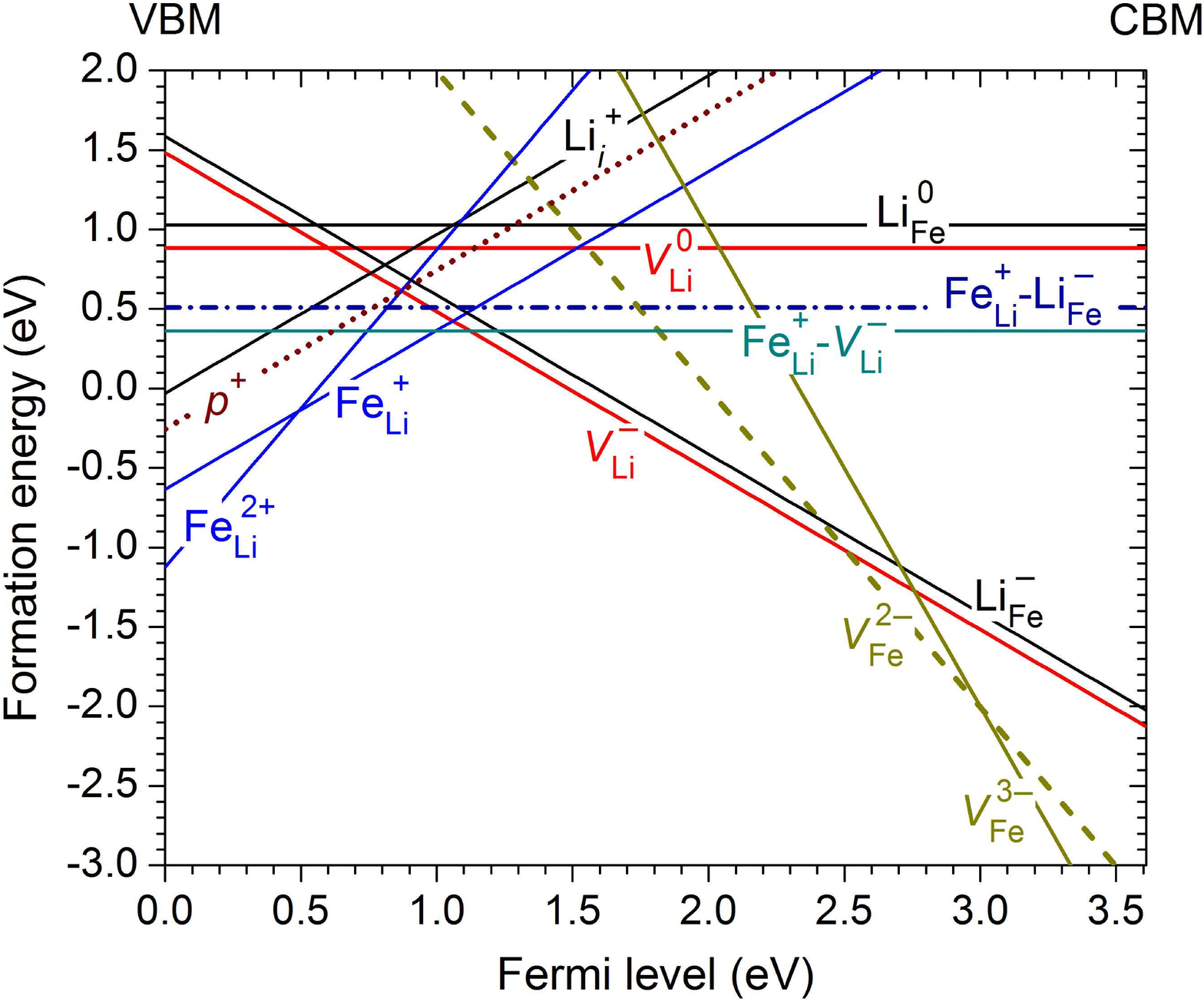}
\end{center}
\vspace{-0.2in}
\caption{Calculated formation energies of native point defects and defect complexes in LiFePO$_{4}$, plotted as a function of Fermi level with respect to the VBM. The energies are obtained at Point A in the chemical-potential diagram for $\mu_{{\rm O}_{2}}$=$-$4.59 eV ({\it cf.}~Fig.~\ref{fig;chempot}), representing equilibrium with Fe$_{2}$O$_{3}$ and Fe$_{3}$(PO$_{4}$)$_{2}$.}\label{fig;459A}
\end{figure}

Figure \ref{fig;459A} shows the calculated formation energies of relevant native point defects and defect complexes in LiFePO$_{4}$ for a representative oxygen chemical potential value, $\mu_{{\rm O}_{2}}$=$-$4.59 eV, and $\mu_{{\rm Li}}$=$-$2.85 eV, $\mu_{{\rm Fe}}$=$-$2.18 eV, and $\mu_{{\rm P}}$=$-$4.64 eV. This set of atomic chemical potentials corresponds to Point A in Fig.~\ref{fig;chempot}, representing the limiting case (Li-deficient) where Fe$_{2}$O$_{3}$, Fe$_{3}$(PO$_{4}$)$_{2}$, and LiFePO$_{4}$ are in equilibrium. The slope in the formation energy plots indicates the charge state. Positive slope indicates that the defect is positively charged, negative slope indicates the defect is negatively charged. With the chosen set of atomic chemical potentials, the positively charged iron antisite Fe$_{\rm Li}^{+}$ and negatively charged lithium vacancy ($V_{\rm Li}^{-}$) have the lowest formation energies among the charged point defects for a wide range of Fermi-level values. While there are different charged point defects coexisting in the system with different concentrations, the ones with the lowest formation energies have the highest concentrations and are dominant.\cite{peles2007,hoang2009,wilson-short} Figure \ref{fig;459A} indicates that, in the absence of electrically active impurities that can affect the Fermi-level position, or when such impurities occur in much lower concentrations than charged native defects, the Fermi level will be pinned at $\epsilon_{F}$=1.06 eV, where the formation energies and hence, approximately, the concentrations of Fe$_{\rm Li}^{+}$ and $V_{\rm Li}^{-}$ are equal. Also, charged native defects have positive formation energies only near $\epsilon_{F}$=1.06 eV. Therefore, any attempt to deliberately shift the Fermi level far away from this position and closer to the VBM or CBM, e.g., via doping with acceptors or donors, will result in positively or negatively charged native defects having negative formation energies, i.e., the native defects will form spontaneously and counteract the effects of doping.\cite{walle:3851,janotti2009,vdwalle2010,catlow2011} This indicates that LiFePO$_{4}$ cannot be doped $p$-type or $n$-type. In the following, we analyze in detail the structure and energetics of the native defects. The dependence of defect formation energies on the choice of atomic chemical potentials will be discussed in the next section.

{\bf Small Hole Polarons.} The creation of a {\it free} positively charged (hole) polaron $p^{+}$ (i.e., $p^{+}$ in the absence of other defects or extrinsic impurities) involves removing one electron from the LiFePO$_{4}$ supercell (hereafter referred to as ``the system''). This results in the formation of a Fe$^{3+}$ site in the system. The calculated magnetic moment at this (Fe$^{3+}$) site is 4.28 $\mu_{\rm B}$, compared to 3.76 $\mu_{\rm B}$ at other iron (Fe$^{2+}$) sites. The local geometry near the Fe$^{3+}$ site is slightly distorted with the neighboring O atoms moving toward Fe$^{3+}$; the average Fe-O bond length is 2.07 {\AA}, compared to 2.18 {\AA} of the other Fe-O bonds. Note that in pristine FePO$_{4}$, the delithiated phase of LiFePO$_{4}$, the calculated magnetic moment is 4.29 $\mu_{\rm B}$ at the iron (Fe$^{3+}$) sites, and the calculated average Fe-O bond length is 2.06 {\AA}. This indicates that a hole (created by removing an electron from the system) has been successfully stabilized at one of the iron sites and the lattice geometry is locally distorted, giving rise to a hole polaron in LiFePO$_{4}$. Since the local distortion is found to be mostly limited to the neighboring O atoms of the Fe$^{3+}$ site, this hole polaron is considered as small polaron where the hole is ``self-trapped'' in its own potential.\cite{Shluger1993,Stoneham2007} The formation of {\it free} hole polarons in LiFePO$_{4}$ is necessarily related to the rather strong interaction between Fe 3$d$ and O $p$ states, and the fact that the VBM consists predominantly of the highly localized $d$ states.

We have investigated the migration path of $p^{+}$ and estimated the energy barrier using the NEB method.\cite{ci-neb} The migration of $p^{+}$ involves an electron and its associated lattice distortion being transferred from a Fe$^{2+}$ site to a neighboring Fe$^{3+}$ site. Since spin conservation is required in this process, we carried out our calculations not using the ground-state AFM structure of LiFePO$_{4}$ but the FM one where all the spins are aligned in the same direction. We calculated the migration path by sampling the atomic positions between ground-state configurations. For those configurations other than ground-state ones, the atomic positions were kept fixed and only electron density was relaxed self-consistently, similar to the method presented in Ref.\cite{Maxisch:2006p103}. The migration barrier is the energy difference between the highest-energy configuration and the ground state. We find that the migration barrier of $p^{+}$ is 0.25 eV between the two nearest Fe sites approximately in the $b$-$c$ plane, which is comparable to that (0.22 eV) reported by in Ref.\cite{Maxisch:2006p103}.

\begin{figure*}
\begin{center}
\includegraphics[width=6.0in]{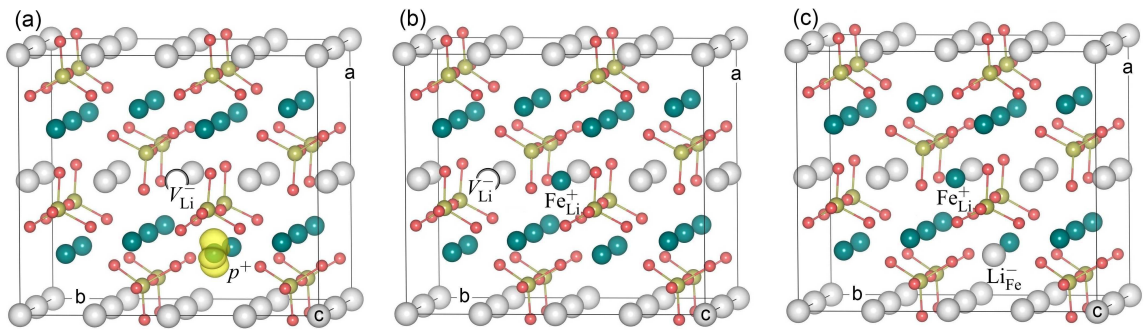}
\end{center}
\vspace{-0.15in}
\caption{Defects in LiFePO$_{4}$: (a) $V_{\rm Li}^{0}$ can be regarded as a complex of $V_{\rm Li}^{-}$ (represented by the empty sphere) and hole polaron $p^{+}$ (i.e., Fe$^{3+}$; decorated with the square of the wavefunctions of the lowest unoccupied state in the electronic structure of LiFePO$_{4}$ in the presence of $V_{\rm Li}^{0}$); (b) Fe$_{\rm Li}^{+}$-$V_{\rm Li}^{-}$, a complex of Fe$_{\rm Li}^{+}$ and $V_{\rm Li}^{-}$; and (c) Fe$_{\rm Li}^{+}$-Li$_{\rm Fe}^{-}$, a complex of Fe$_{\rm Li}^{+}$ and Li$_{\rm Fe}^{-}$. Large (gray) spheres are Li, medium (blue) spheres Fe, small (yellow) spheres P, and smaller (red) spheres O.}\label{fig;defect_structure}
\end{figure*}

{\bf Vacancies and Interstitials.} Negatively charged lithium vacancies ($V_{\rm Li}^{-}$) are created by removing a Li$^{+}$ ion from the system. Since, in LiFePO$_{4}$, Li donates one electron to the lattice one expects that the removal of Li$^{+}$ causes only a small disturbance in the system. Indeed we see that lattice relaxations around the void formed by the removed Li$^{+}$ are negligible. The energy needed to form $V_{\rm Li}^{-}$ should also be small, consistent with our results in Fig.~\ref{fig;459A}. $V_{\rm Li}^{0}$, on the other hand, is created by removing a Li atom (i.e., Li$^{+}$ and an electron) from the system. This leads to the formation of a void (at the site of the removed Li$^{+}$) and an Fe$^{3+}$ (formed by the removed electron) at the neighboring Fe site. Similar to the {\it free} hole polaron, the neighboring O atoms of the Fe$^{3+}$ site in $V_{\rm Li}^{0}$ also move toward Fe$^{3+}$, with the average Fe-O distance being 2.07 {\AA}. The calculated magnetic moment is 4.29 $\mu_{\rm B}$ at the Fe$^{3+}$ site, equal to that at the Fe$^{3+}$ site in the case of a free polaron. $V_{\rm Li}^{0}$, therefore, should be regarded as a complex of $V_{\rm Li}^{-}$ and $p^{+}$, with the two defects being 3.26 {\AA} apart. Figure \ref{fig;defect_structure}(a) shows the structure of $V_{\rm Li}^{0}$. The binding energy of $V_{\rm Li}^{0}$ is 0.34 eV (with respect to $V_{\rm Li}^{-}$ and $p^{+}$). Note that this value is 0.42 eV in our calculations using (1$\times$3$\times$3) supercells which have 252 atoms/cell. Our estimated binding energy is thus comparable to that of 0.39 and about 0.50 eV reported by Fisher et al.\cite{Fisher:2008p80} and Maxisch et al.,\cite{Maxisch:2006p103} respectively. For lithium interstitials, the stable defect is Li$_{i}^{+}$, created by adding Li$^{+}$ into the system. Other charge states of $V_{\rm Li}$ and Li$_{i}$ are not included in Fig.~\ref{fig;459A} because they either have too high energies to be relevant or are unstable.

The migration path of $V_{\rm Li}^{-}$ is calculated by moving a Li$^{+}$ unit from a nearby lattice site into the vacancy. The energy barrier for $V_{\rm Li}^{-}$ is estimated to be 0.32 eV along the $b$-axis and 2.27 eV along the $c$-axis. This suggests that, in the absence of other native defects and extrinsic impurities, lithium diffusion in LiFePO$_{4}$ is highly one-dimensional along the Li channels ($b$-axis) because the energy barrier to cross between the channels is too high. The migration path of $V_{\rm Li}^{-}$ is, however, not a straight line but a curved path along the $b$-axis. Our results are thus in general agreement with previously reported theoretical studies\cite{morgan:A30,Islam:2005p168,Fisher:2008p80,Adams:2010p132} and experimental observation.\cite{Nishimura2008} The estimated energy barriers for the migration of $V_{\rm Li}^{-}$ along the $b$ and $c$ axes are lower than those (0.55 and 2.89 eV, respectively) reported by Islam et al.\cite{Islam:2005p168} obtained from calculations using inter-atomic potentials, but closer to those (0.27 and about 2.50 eV) reported by Morgan et al.\cite{morgan:A30} obtained in GGA calculations with smaller supercells. For $V_{\rm Li}^{0}$, a complex of $p^{+}$ and $V_{\rm Li}^{-}$, one can estimate the lower bound of the migration barrier by taking the higher of the migration energies of the constituents,\cite{wilson-short} which is 0.32 eV (along the $b$-axis), the value for $V_{\rm Li}^{-}$.

Other possible vacancies in LiFePO$_{4}$ are those associated with Fe$^{2+}$ and (PO$_{4}$)$^{3-}$ units. The creation of $V_{\rm Fe}^{2-}$ corresponds to removing Fe$^{2+}$ from the system. We find that this negatively charged defect causes significant relaxations in the lattice geometry. The neighboring Li$^{+}$ ions move toward the defect, resulting in the Li channels being bent near $V_{\rm Fe}^{2-}$ where Li$^{+}$ ions are displaced up to 0.27 {\AA} from their original positions. $V_{\rm Fe}^{-}$ can be considered as a complex of $V_{\rm Fe}^{2-}$ and $p^{+}$ with the distance between the two defects being 3.81 {\AA}. $V_{\rm Fe}^{3-}$, on the other hand, corresponds to removing Fe$^{2+}$ but leaving an electron in the system. This defect can be regarded as a complex of $V_{\rm Fe}^{2-}$ and a negatively charged (electron) polaron (hereafter denoted as $p^{-}$). At the Fe site where the electron polaron resides, which is 7.68 {\AA} from the vacancy, the calculated magnetic moment is 2.86 $\mu_{\rm B}$; the average Fe-O distance is 2.30 {\AA}, which is larger than that associated with other Fe sites (2.18 {\AA}). Finally, $V_{{\rm PO}_{4}}$ is stable as $V_{{\rm PO}_{4}}^{3+}$ as expected. This positively charged defect corresponds to removing the whole (PO$_{4}$)$^{3-}$ unit from the system. With the chosen set of atomic chemical potentials, $V_{\rm Fe}^{-}$ and $V_{{\rm PO}_{4}}^{3+}$ have very high formation energies (2.33 eV and 3.56 eV, respectively, at $\epsilon_{F}$=1.06 eV) and are therefore not included in Fig.~\ref{fig;459A}.

{\bf Antisite Defects.} Lithium antisites Li$_{\rm Fe}$ are created by replacing Fe at an Fe site with Li. Li$_{\rm Fe}^{-}$ can be considered as replacing Fe$^{2+}$ with Li$^{+}$. Due to the Coulombic interaction, the two nearest Li$^{+}$ ion neighbors of Li$_{\rm Fe}^{-}$ are pulled closer to the negatively charged defect with the distance being 3.25 {\AA}, compared to 3.32 {\AA} of the equivalent bond in pristine LiFePO$_{4}$. Li$_{\rm Fe}^{0}$, on the other hand, can be regarded as a complex of Li$_{\rm Fe}^{-}$ and $p^{+}$ with the distance between the two defects being 3.98 {\AA}. The binding energy of Li$_{\rm Fe}^{0}$ (with respect to Li$_{\rm Fe}^{-}$ and $p^{+}$) is 0.30 eV. Similarly, one can replace Li at an Li site with Fe, which creates an iron antisite Fe$_{\rm Li}$. Fe$_{\rm Li}^{+}$ corresponds to replacing Li$^{+}$ with Fe$^{2+}$, whereas Fe$_{\rm Li}^{2+}$ can be regarded as a complex of Fe$_{\rm Li}^{+}$ and $p^{+}$. For Fe$_{\rm Li}^{0}$, which corresponds to replacing one Li$^{+}$ with Fe$^{2+}$ and adding an extra electron to the system, the extra electron is stabilized at the substituting Fe atom, where the calculated magnetic moment is 2.95 $\mu_{\rm B}$. One might also regard Fe$_{\rm Li}^{0}$ as a complex of Fe$_{\rm Li}^{+}$ and $p^{-}$, but in this case the two defects stay at the same lattice site. With the chosen set of chemical potentials, Fe$_{\rm Li}^{0}$ has a very high formation energy (2.04 eV) and is therefore not included in Fig.~\ref{fig;459A}. Again, other native defects that are not included here are unstable or have too high formation energies to be relevant.

{\bf Defect Complexes.} From the above analyses, it is clear that defects such as $p^{+}$ ($p^{-}$), $V_{\rm Li}^{-}$, $V_{\rm Fe}^{2-}$, Fe$_{\rm Li}^{+}$, Li$_{\rm Fe}^{-}$, and $V_{{\rm PO}_{4}}^{3+}$ can be considered as {\it elementary} native defects in LiFePO$_{4}$, i.e., the structure and energetics of other native defects can be interpreted in terms of these basic building blocks. This is similar to what has been observed in complex hydrides.\cite{wilson-short} These elementary defects (except the free polarons) are, in fact, point defects that are formed by adding and/or removing {\it only} Li$^{+}$, Fe$^{2+}$, and (PO$_{4}$)$^{3-}$ units. They have low formation energies ({\it cf.}~Fig.~\ref{fig;459A}) because the addition/removal of these units causes the least disturbance to the system, which is consistent with the simple bonding picture for LiFePO$_{4}$ presented in the previous section. The identification of the elementary native defects, therefore, not only helps us gain a deeper understanding of the structure and energetics of the defects in LiFePO$_{4}$ but also has important implications. For example, one should treat the migration of defects such as $V_{\rm Li}^{0}$ as that of a $V_{\rm Li}^{-}$ and $p^{+}$ complex with a finite binding energy, rather than as a single point defect.

In addition to the defect complexes that involve $p^{+}$ and $p^{-}$ such as $V_{\rm Li}^{0}$, $V_{\rm Fe}^{-}$, $V_{\rm Fe}^{3-}$, Li$_{\rm Fe}^{0}$, Fe$_{\rm Li}^{0}$, and Fe$_{\rm Li}^{2+}$ described above, we also considered those consisting of $V_{\rm Li}^{-}$, Fe$_{\rm Li}^{+}$, Li$_{\rm Fe}^{-}$, and $V_{\rm Fe}^{2-}$ such as Fe$_{\rm Li}^{+}$-Li$_{\rm Fe}^{-}$, Fe$_{\rm Li}^{+}$-$V_{\rm Li}^{-}$, and 2Fe$_{\rm Li}^{+}$-$V_{\rm Fe}^{2-}$. Figure \ref{fig;defect_structure}(b) shows the structure of Fe$_{\rm Li}^{+}$-$V_{\rm Li}^{-}$. The distance between Fe$_{\rm Li}^{+}$ and $V_{\rm Li}^{-}$ is 2.96 {\AA} (along the $b$-axis), compared to 3.03 {\AA} between the two Li sites in pristine LiFePO$_{4}$. We find that this complex has a formation energy of 0.36$-$0.56 eV for reasonable choices of atomic chemical potentials, and a binding energy of 0.49 eV. With such a relatively high binding energy, even higher than the formation energy of isolated Fe$_{\rm Li}^{+}$ and $V_{\rm Li}^{-}$ (0.42 eV at $\epsilon_{F}$=1.06 eV, {\it cf.}~Fig.~\ref{fig;459A}), Fe$_{\rm Li}^{+}$-$V_{\rm Li}^{-}$ is expected to occur with a concentration larger than either of its constituents under thermal equilibrium conditions during synthesis.\cite{walle:3851} In Fe$_{\rm Li}^{+}$-$V_{\rm Li}^{-}$, the energy barrier for migrating Fe$_{\rm Li}^{+}$ to $V_{\rm Li}^{-}$ is about 0.74 eV, comparable to that (0.70 eV) reported by Fisher et al.\cite{Fisher:2008p80} This value is twice as high as the migration barrier of $V_{\rm Li}^{-}$, indicating Fe$_{\rm Li}^{+}$ has low mobility.

Figure \ref{fig;defect_structure}(c) shows the structure of Fe$_{\rm Li}^{+}$-Li$_{\rm Fe}^{-}$. This antisite pair has a formation energy of 0.51 eV. This value is independent of the choice of chemical potentials because the chemical potential term in the formation energy formula cancels out, {\it cf.}~Eq.~(\ref{eq;eform}). Fe$_{\rm Li}^{+}$-Li$_{\rm Fe}^{-}$ has a binding energy of 0.44 eV; the distance between Fe$_{\rm Li}^{+}$ and Li$_{\rm Fe}^{-}$ is 3.45 {\AA}, compared to 3.32 {\AA} between the lithium and iron sites. Finally, we find that 2Fe$_{\rm Li}^{+}$-$V_{\rm Fe}^{2-}$ has a formation energy of 1.47$-$1.67 eV for reasonable choices of the atomic chemical potentials, and a binding energy of 1.25 eV. With this high formation energy, the complex is unlikely to form in LiFePO$_{4}$, and is therefore not included in Fig.~\ref{fig;459A}. Note that the formation energies of Fe$_{\rm Li}^{+}$-$V_{\rm Li}^{-}$ and 2Fe$_{\rm Li}^{+}$-$V_{\rm Fe}^{2-}$ have the same dependence on the atomic chemical potentials (both contain the term $-\mu_{\rm Fe}+2\mu_{\rm Li}$) and, hence, the same dependence on $\mu_{{\rm O}_{2}}$. For {\it any} given set of chemical potentials, the formation energy of 2Fe$_{\rm Li}^{+}$-$V_{\rm Fe}^{2-}$ is higher than that of Fe$_{\rm Li}^{+}$-$V_{\rm Li}^{-}$ by 1.11 eV. We also considered possible lithium and iron Frenkel pairs (i.e., interstitial-vacancy pairs), but these pairs are unstable toward recombination, probably because there is no energy barrier or too small of a barrier between the vacancy and the interstitial.

The above mentioned neutral defect complexes have also been studied by other research groups using either interatomic-potential simulations\cite{Islam:2005p168,Fisher:2008p80} or first-principles DFT calculations.\cite{Malik2010} Islam et al.~found that Fe$_{\rm Li}^{+}$-Li$_{\rm Fe}^{-}$ has a formation energy of 0.74 eV (or 1.13 eV if the two defects in the pair are considered as isolated defects) and a binding energy of 0.40 eV, and is energetically most favorable among possible native defects.\cite{Islam:2005p168} The reported formation energy is, however, higher than our calculated value by 0.23 eV. This difference may be due to the different methods used in the calculations. Fisher et al.~reported a formation energy of 3.13 eV for Fe$_{\rm Li}^{+}$-$V_{\rm Li}^{-}$,\cite{Fisher:2008p80} which is much higher than our calculated value. Note, however, that Fisher et al.~assumed the reaction ${\rm FeO}+2{\rm Li}_{\rm {Li}}^{0}$$\rightarrow$${\rm Fe}_{\rm Li}^{+}+V_{\rm Li}^{-}+{\rm Li}_{2}{\rm O}$ for the formation of Fe$_{\rm Li}^{+}$-$V_{\rm Li}^{-}$ which implies that LiFePO$_{4}$ is in equilibrium with FeO and Li$_{2}$O. This scenario is unlikely to occur, as indicated in the Li-Fe-P-O$_{2}$ phase diagrams calculated by Ong et al.,\cite{ong2008} where equilibrium between these phases has never been observed. This may also be the reason that the formation energy of $V_{\rm Li}^{0}$ reported by the same authors (4.41 eV)\cite{Fisher:2008p80} is much higher than our calculated values.

Based on first-principles calculations, Malik et al.~reported a formation energy of 0.515$-$0.550 eV for the antisite pair Fe$_{\rm Li}^{+}$-Li$_{\rm Fe}^{-}$,\cite{Malik2010} which is very close to our calculated value (0.51 eV). For 2Fe$_{\rm Li}^{+}$-$V_{\rm Fe}^{2-}$, the formation energy was reported to be of about 1.60$-$1.70 eV for $\mu_{{\rm O}_{2}}$ ranging from $-$3.03 to $-$8.21 eV,\cite{Malik2010} which is also comparable to our results. Malik et al., however, obtained a much higher formation energy for Fe$_{\rm Li}^{+}$-$V_{\rm Li}^{-}$, from about 3.60 to 5.10 eV for the same range of $\mu_{{\rm O}_{2}}$ values.\cite{Malik2010} This energy is much higher than that obtained in our calculations (0.36$-$0.56 eV). Although we have no explanation for this discrepancy, we observe that the calculated formation energies of 2Fe$_{\rm Li}^{+}$-$V_{\rm Fe}^{2-}$ and Fe$_{\rm Li}^{+}$-$V_{\rm Li}^{-}$ in Malik et al.'s work have distinct $\mu_{{\rm O}_{2}}$-dependencies (see Figure S1 in the Supporting Information of Ref.~\cite{Malik2010}), instead of having the same dependence on $\mu_{{\rm O}_{2}}$ as we discussed above, indicating their scheme of accounting for the atomic chemical potentials differs from the standard procedure.

\section{\label{sec;tuning}Tailoring Defect Concentrations}

It is important to note that the {\it energy landscape} presented in Fig.~\ref{fig;459A} may change as one changes the atomic chemical potentials, i.e., synthesis conditions. The calculated formation energies are a function of four variables $\mu_{\rm Li}$, $\mu_{\rm Fe}$, $\mu_{\rm P}$, and $\mu_{{\rm O}_{2}}$, which in turn depend on each other and vary within the established constraints. A change in one variable leads to changes in the other three. In the following discussions, we focus on two ``knobs'' that can be used to experimentally tailor the formation energy and hence the concentration of different native defects in LiFePO$_{4}$, and suppress or enhance certain defects for targeted applications. One is $\mu_{{\rm O}_{2}}$, which can be controlled by controlling temperature and pressure and/or oxygen reducing agents. Lower $\mu_{{\rm O}_{2}}$ values represent the so-called ``more reducing environments,'' which are usually associated with higher temperatures and/or lower oxygen partial pressures and/or the presence of oxygen reducing agents; whereas higher $\mu_{{\rm O}_{2}}$ values represent ``less reducing environments.''\cite{ong2008} The other is the degree of lithium off-stoichiometry with respect to LiFePO$_{4}$ exhibited through the tendency toward formation of Li-containing or Fe-containing secondary phases in the synthesis of LiFePO$_{4}$. As discussed previously, in the environments to which we refer as Li-excess (Li-deficient), the system is close to forming Li-containing (Fe-containing) secondary phases.

\begin{figure} 
\begin{center} 
\includegraphics[width=3.0in]{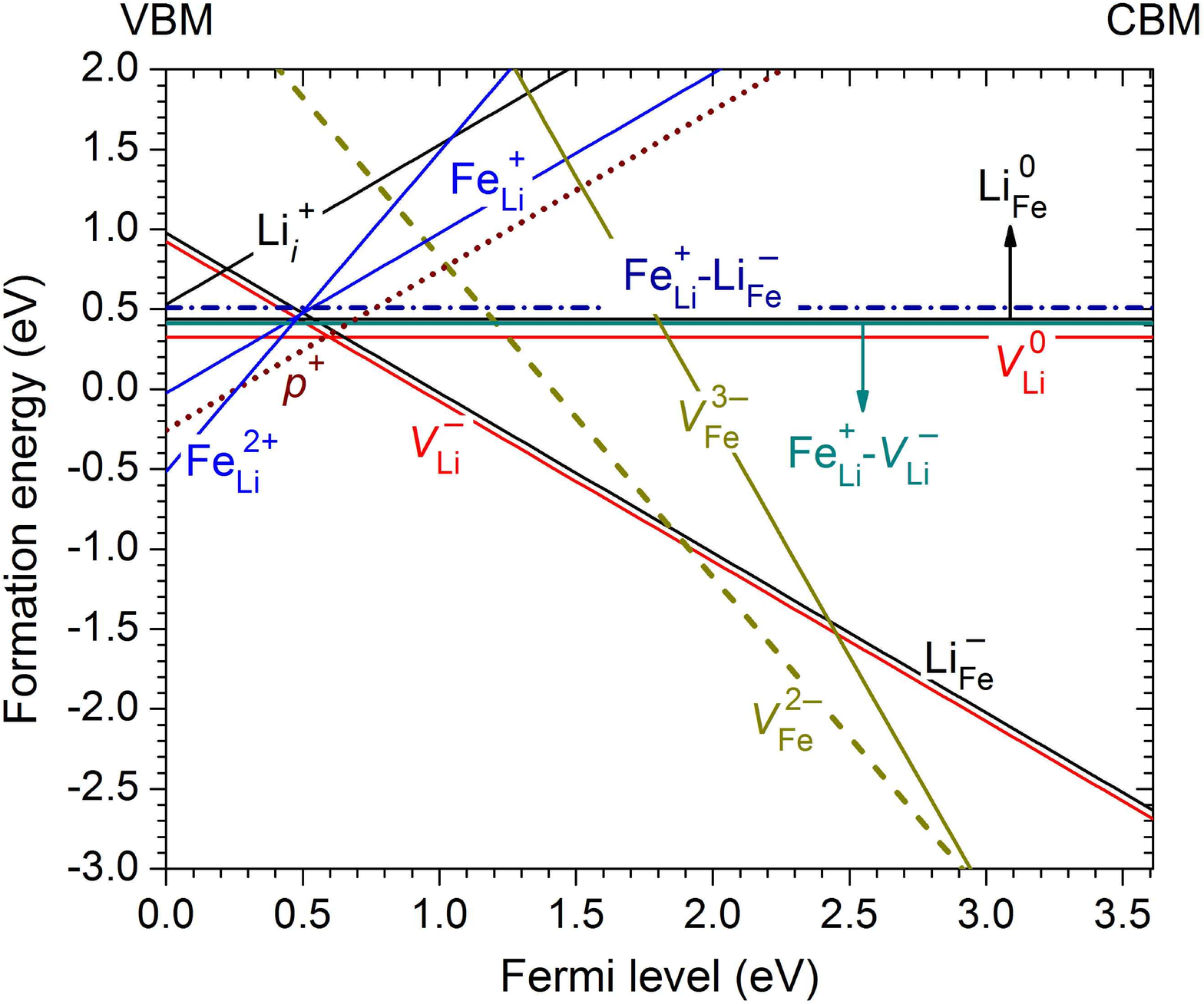} 
\end{center} 
\vspace{-0.15in} 
\caption{Calculated formation energies of 
native point defects and defect complexes in LiFePO$_{4}$, plotted as a function of Fermi level with respect to the VBM. The energies are obtained at $\mu_{{\rm O}_{2}}$=$-$3.03 eV, and equilibrium with Fe$_{2}$O$_{3}$ and Fe$_{7}$(PO$_{4}$)$_{6}$ is assumed.}\label{fig;303A} 
\end{figure} 

{\bf Varying the Atomic Chemical Potentials.} Let us assume, for example, Li-deficient environments and vary $\mu_{{\rm O}_{2}}$ from $-$3.03 (where LiFePO$_{4}$ first starts to form) to $-$8.25 eV (where it ceases to form).\cite{ong2008} This amounts to choosing different cuts along the $\mu_{{\rm O}_{2}}$ axis in Fig.~\ref{fig;chempot} to give different two-dimensional polygons of LiFePO$_4$ stability. Figure \ref{fig;303A} shows the calculated formation energies for $\mu_{{\rm O}_{2}}$=$-$3.03 eV, assuming equilibrium with Fe$_{2}$O$_{3}$ and Fe$_{7}$(PO$_{4}$)$_{6}$ (i.e., Li-deficient) which gives rise to $\mu_{\rm Li}$=$-$3.41, $\mu_{\rm Fe}$=$-$3.35, and $\mu_{\rm P}$=$-$6.03 eV. Figure \ref{fig;303A} clearly shows changes in the energy landscape of the defects, compared to Fig.~\ref{fig;459A}. The lowest energy point defects that determine the Fermi-level position are now $p^{+}$ and $V_{\rm Li}^{-}$.  Near $\epsilon_{F}$=0.59 eV where $p^{+}$ and $V_{\rm Li}^{-}$ have equal formation energies, $V_{\rm Li}^{0}$ also has the lowest energy. This indicates that, under high $\mu_{{\rm O}_{2}}$ and Li-deficient environments, $p^{+}$ and $V_{\rm Li}^{-}$ are the dominant native point defects in LiFePO$_{4}$ and are likely to exist in the form of the neutral complex $V_{\rm Li}^{0}$. Note that, with the chosen set of atomic chemical potentials, Li$_{\rm Fe}^{-}$ also has a low formation energy, very close to that of $V_{\rm Li}^{-}$, indicating the presence of a relatively high concentration of Li$_{\rm Fe}^{-}$. Similar to $V_{\rm Li}^{-}$, Li$_{\rm Fe}^{-}$ can combine with $p^{+}$ to form Li$_{\rm Fe}^{0}$. However, since Li$_{\rm Fe}^{0}$ has a higher formation energy and a smaller binding energy than $V_{\rm Li}^{0}$ , only a small portion of Li$_{\rm Fe}^{-}$ is expected to be stable in form of Li$_{\rm Fe}^{0}$ under thermal equilibrium conditions. Iron vacancies have the lowest energies in a wide range of the Fermi-level values as expected, given the very low iron chemical potential.

\begin{figure}
\begin{center}
\includegraphics[width=3.0in]{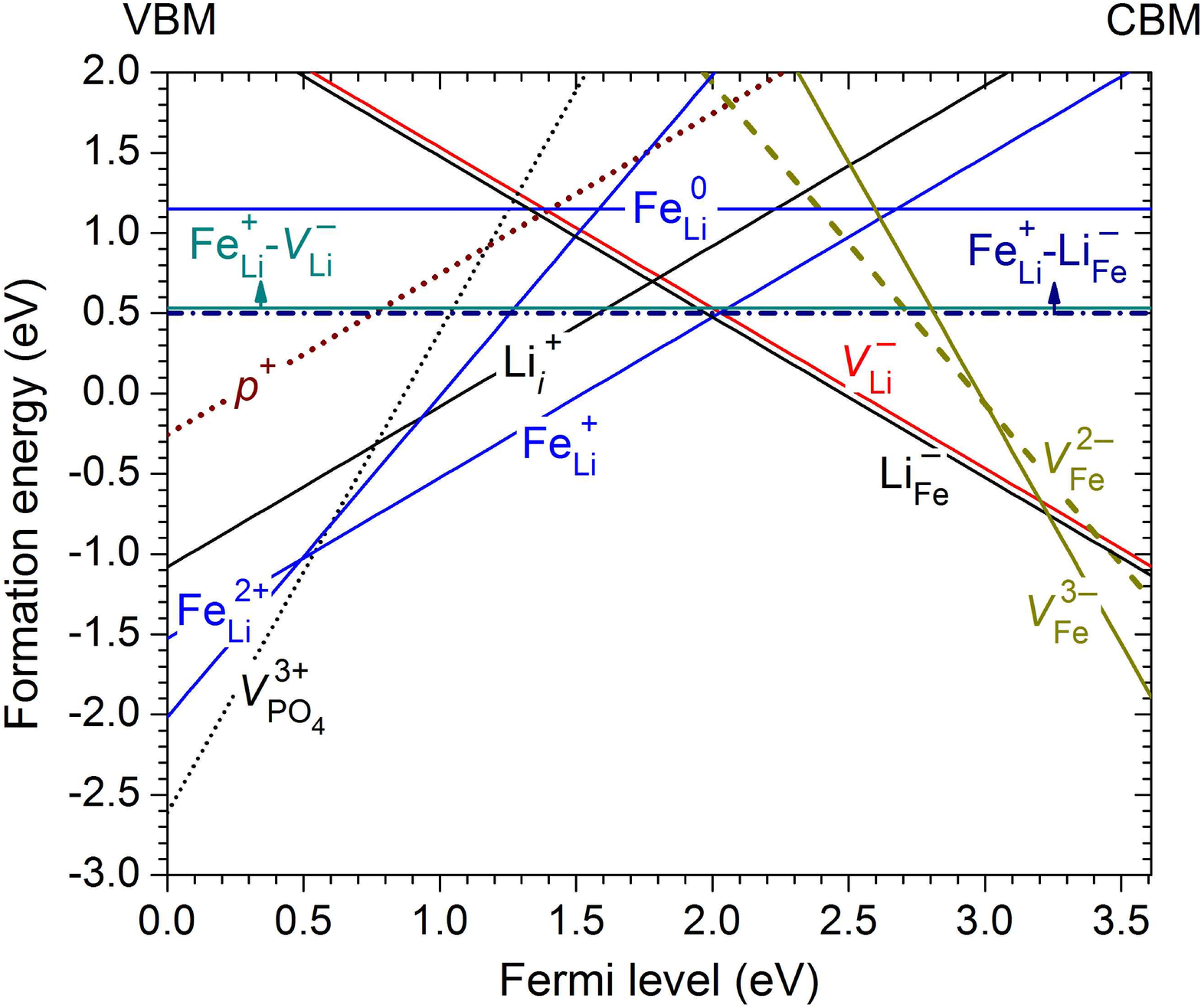}
\end{center}
\vspace{-0.15in}
\caption{Calculated formation energies of native point defects and defect complexes in LiFePO$_{4}$, plotted as a function of Fermi level with respect to the VBM. The energies are obtained at $\mu_{{\rm O}_{2}}$=$-$8.21 eV, and equilibrium with Fe$_{2}$P and Fe$_{3}$P is assumed.}\label{fig;821A}
\end{figure}

Figure \ref{fig;821A} shows the calculated formation energies for $\mu_{{\rm O}_{2}}$=$-$8.21 eV. The formation energies are obtained by assuming equilibrium with Fe$_{2}$P and Fe$_{3}$P (i.e., Li-deficient) which gives rise to $\mu_{\rm Li}$=$-$1.80, $\mu_{\rm Fe}$=$-$0.24, and $\mu_{\rm P}$=$-$0.39 eV. We find that Fe$_{\rm Li}^{+}$ and Li$_{\rm{Fe}}^{-}$ are now the dominant native point defects, pinning the Fermi level at $\epsilon_{F}$=2.00 eV. The complex Fe$_{\rm Li}^{+}$-Li$_{\rm{Fe}}^{-}$ has a binding energy of 0.44 eV, comparable to the formation energies of Fe$_{\rm Li}^{+}$ and Li$_{\rm{Fe}}^{-}$ (which are both 0.48 eV at $\epsilon_{F}$=2.00 eV). This suggests that Fe$_{\rm Li}^{+}$ and Li$_{\rm{Fe}}^{-}$ are likely to exist both in the form of Fe$_{\rm Li}^{+}$-Li$_{\rm{Fe}}^{-}$, but also as isolated point defects. With this set of atomic chemical potentials, we find that $V_{{\rm PO}_{4}}^{3+}$ has the lowest formation energy near the VBM, and Fe$_{\rm Li}^{0}$ has a formation energy of 1.15 eV that, while very high, is lower than $V_{\rm Li}^{0}$ ($1.93$ eV) and Li$_{\rm Fe}^{0}$ ($1.98$ eV), which were found to have lower formation energies under different conditions ({\it cf.}~Figures 3 and 5).

We also investigated the dependence of defect formation energies on $\mu_{\rm {Li}}$ (and $\mu_{\rm {Fe}}$), i.e., Li-deficiency versus Li-excess, for a given $\mu_{{\rm O}_{2}}$ value. For $\mu_{{\rm O}_{2}}$=$-4.59$ eV, for example, the results obtained at Points B and C in Fig.~\ref{fig;chempot} show energy landscapes that are similar to that at Point A (Li-deficient), namely Fe$_{\rm Li}^{+}$ and $V_{\rm Li}^{-}$ are the dominant native point defects in LiFePO$_{4}$ and likely to exist in form of Fe$_{\rm Li}^{+}$-$V_{\rm Li}^{-}$. At Point D, where LiFePO$_{4}$ is in equilibrium with Li$_{4}$P$_{2}$O$_{7}$ and Li$_{3}$PO$_{4}$ (Li-excess), we find instead that Fe$_{\rm Li}^{+}$ and Li$_{\rm Fe}^{-}$ are energetically most favorable, and are likely to exist as Fe$_{\rm Li}^{+}$-Li$_{\rm{Fe}}^{-}$. The calculated formation energy of $p^{+}$ is only slightly higher than that of Fe$_{\rm Li}^{+}$, indicating a coexisting high concentration of $p^{+}$. The hole polarons in this case are expected to exist as isolated defects under thermal equilibrium since Li$_{\rm Fe}^{0}$ has a relatively high formation energy (0.66 eV) and a small binding energy (0.30 eV). Point E gives results that are similar to those at Point D. In contrast, when we choose $\mu_{{\rm O}_{2}}$=$-8.21$ eV, we find that Fe$_{\rm Li}^{+}$ and Li$_{\rm Fe}^{-}$ are the most energetically favorable defects regardless of the choice of phase-equilibrium conditions.

\begin{table*}
\caption{Calculated formation energies ($E^{f}$) and migration barriers ($E_{m}$) of the most relevant native point defects and defect complexes in LiFePO$_4$. (1)$-$(8) are the equilibrium conditions; see text. Binding energies ($E_{b}$) of the defect complexes (with respect to their isolated constituents) are given in the last column. The formation energy of 2Fe$_{\rm{Li}}^+$-$V_{\rm{Fe}}^{2-}$ is high and thus the complex is not likely to form, but is also given here for comparison.}\label{tab}
\begin{ruledtabular}
\begin{tabular}{lccccccccllc}
Defect&\multicolumn{8}{c}{$E^{f}$ (eV)}&  $E_{m}$ (eV)&Constituents&$E_{b}$(eV) \\
&(1)&(2)&(3)&(4)&(5)&(6)&(7)&(8)&&& \\
\hline
$p^{+}$&0.33&0.32&0.54&0.40&0.80&0.49&1.74&1.72&0.25& \\
Fe$_{\rm{Li}}^+$&0.57&0.63&0.43&0.55&0.42&0.48&0.48&0.48&&\\
$V_{\rm{Li}}^-$&0.33&0.41&0.43&0.47&0.42&0.56&0.53&0.55&0.32&	\\
Li$_{\rm{Fe}}^-$&0.39&0.32&0.52&0.40&0.53&0.48&0.48&0.48&&\\
$V_{\rm{Li}}^0$&0.32&0.38&0.62&0.52&0.88&0.70&1.93&1.95&0.32\textsuperscript{\emph{a}}&$V_{\rm{Li}}^-$ + $p^{+}$&0.34, 0.42\textsuperscript{\emph{b}}\\
Li$_{\rm{Fe}}^0$&0.42&0.34&0.76&0.50&1.03&0.66&1.92&1.90&&Li$_{\rm{Fe}}^-$ + $p^{+}$&0.30\\
Fe$_{\rm{Li}}^+$-$V_{\rm{Li}}^-$&0.41&0.55&0.37&0.53&0.36&0.55&0.52&0.56&&Fe$_{\rm{Li}}^+$ + $V_{\rm{Li}}^-$&0.49\\
Fe$_{\rm{Li}}^+$-Li$_{\rm{Fe}}^-$&0.51&0.51&0.51&0.51&0.51&0.51&0.51&0.51&&Fe$_{\rm{Li}}^+$ + Li$_{\rm{Fe}}^-$&0.44\\
2Fe$_{\rm{Li}}^+$-$V_{\rm{Fe}}^{2-}$&1.52&1.66&1.48&1.64&1.47&1.66&1.63&1.67&&2Fe$_{\rm{Li}}^+$ + $V_{\rm{Fe}}^{2-}$&1.25\\
\end{tabular}
\textsuperscript{\emph{a}} Lower bound, estimated by considering $V_{\rm{Li}}^0$ as a complex of $V_{\rm{Li}}^-$ and $p^{+}$ and taking the higher of the migration energies of the constituents.
\textsuperscript{\emph{b}} The value obtained in calculations using larger supercells (252 atoms/cell).
\end{ruledtabular}
\end{table*}

{\bf Identifying the General Trends.} We list in Table \ref{tab} the formation energies of the most relevant native point defects and defect complexes in LiFePO$_{4}$, migration barriers of selected point defects, and binding energies of the defect complexes. The chemical potentials are chosen with representative $\mu_{{\rm O}_{2}}$ values and Li-deficient and Li-excess environments to reflect different experimental conditions. Specifically, these conditions represent equilibrium of LiFePO$_{4}$ with (1) Fe$_{2}$O$_{3}$ and Fe$_{7}$(PO$_{4}$)$_{6}$ and (2) Li$_{3}$Fe$_{2}$(PO$_{4}$)$_{3}$ and Li$_{3}$PO$_{4}$, for $\mu_{{\rm O}_{2}}$=$-$3.03 eV; (3) Fe$_{2}$O$_{3}$ and Fe$_{3}$(PO$_{4}$)$_{2}$ and (4) Li$_{4}$P$_{2}$O$_{7}$ and Li$_{3}$PO$_{4}$, for $\mu_{{\rm O}_{2}}$=$-$3.89 eV; (5) Fe$_{2}$O$_{3}$ and Fe$_{3}$(PO$_{4}$)$_{2}$ (i.e., Point A in Fig.~\ref{fig;chempot}) and (6) Li$_{4}$P$_{2}$O$_{7}$ and Li$_{3}$PO$_{4}$ (i.e., Point D in Fig.~\ref{fig;chempot}), for $\mu_{{\rm O}_{2}}$=$-$4.59 eV; (7) Fe$_{3}$P and Fe$_{2}$P and (8) Li$_{3}$PO$_{4}$ and Fe$_{3}$P, for $\mu_{{\rm O}_{2}}$=$-$8.21 eV. Conditions (1), (3), (5), and (7) represent Li-deficient environments, whereas (2), (4), and (6) represent Li-excess. Under each condition, the formation energies for charged defects are taken at the Fermi-level position determined by relevant charged point defects: (1) $\epsilon_{F}$=0.59 eV (where $p^{+}$ and $V_{\rm{Li}}^{-}$ have equal formation energies), (2) $\epsilon_{F}$=0.58 eV ($p^{+}$ and Li$_{\rm{Fe}}^{-}$), (3) $\epsilon_{F}$=0.79 eV (Fe$_{\rm Li}^{+}$ and $V_{\rm{Li}}^{-}$), (4) $\epsilon_{F}$=0.66 eV ($p^{+}$ and Li$_{\rm{Fe}}^{-}$), (5) $\epsilon_{F}$=1.06 eV (Fe$_{\rm Li}^{+}$ and $V_{\rm{Li}}^{-}$), (6) $\epsilon_{F}$=0.74 eV (Fe$_{\rm Li}^{+}$ and Li$_{\rm{Fe}}^{-}$), (7) $\epsilon_{F}$=2.00 eV (Fe$_{\rm Li}^{+}$ and Li$_{\rm{Fe}}^{-}$), and (8) $\epsilon_{F}$=1.98 eV (Fe$_{\rm Li}^{+}$ and Li$_{\rm{Fe}}^{-}$). The results for other $\mu_{{\rm O}_{2}}$ values are not included in Table \ref{tab} because they give results that are similar to those presented here. For example, the energy landscapes for $\mu_{{\rm O}_{2}}$=$-$3.25 eV and $\mu_{{\rm O}_{2}}$=$-7.59$ eV are similar to those for $\mu_{{\rm O}_{2}}$=$-$3.03 eV and $\mu_{{\rm O}_{2}}$=$-4.59$ eV, respectively.

\begin{figure*}[t]
\begin{center}
\includegraphics[width=5.0in]{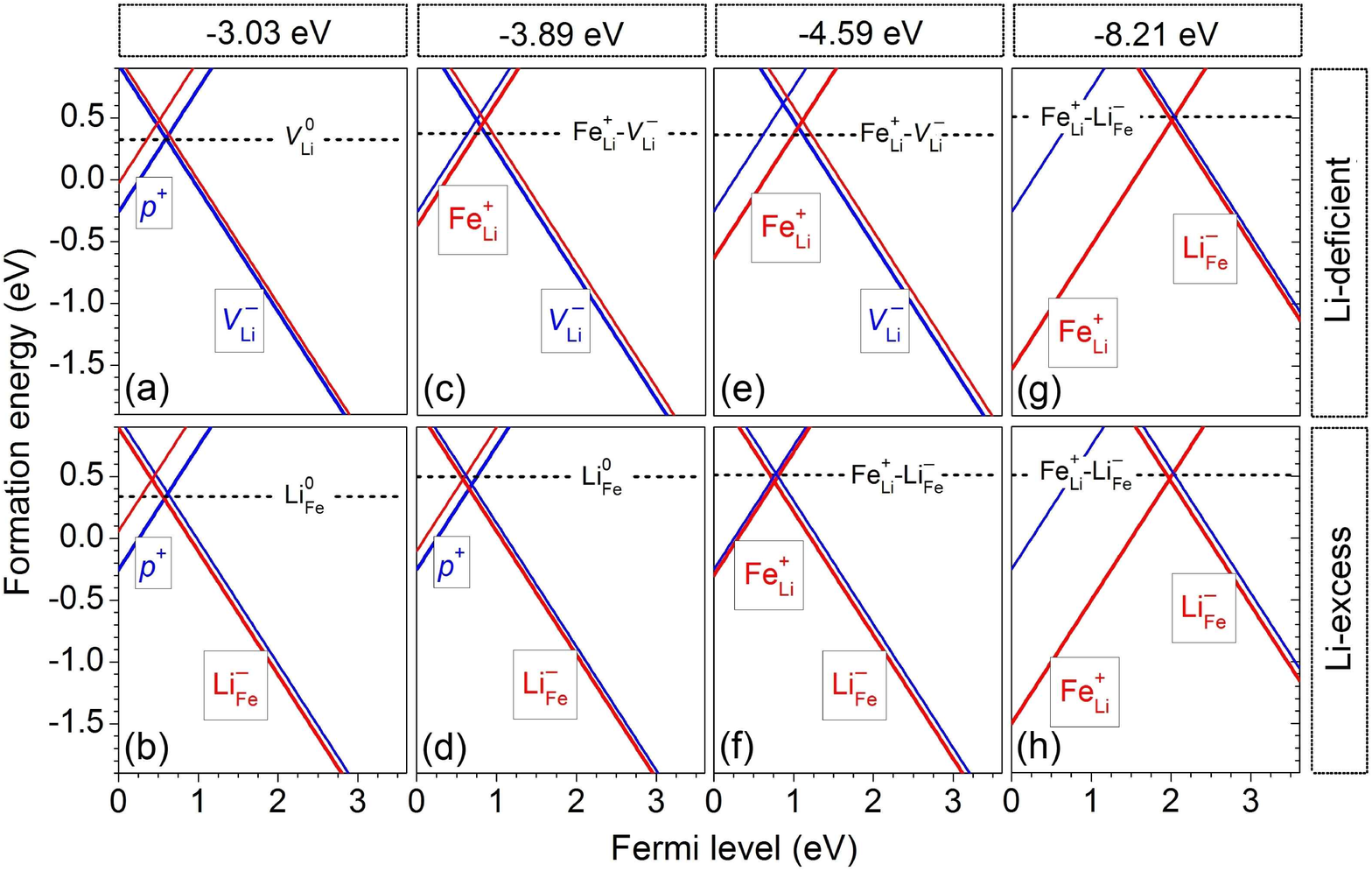}
\end{center}
\vspace{-0.2in}
\caption{Calculated formation energies of the low-energy positively and negatively charged point defects (i.e., $p^{+}$, Fe$_{\rm{Li}}^+$, $V_{\rm{Li}}^-$, and Li$_{\rm{Fe}}^-$) and their neutral defect complexes, plotted as a function of Fermi level with respect to the VBM, under different conditions: $\mu_{{\rm O}_{2}}$=$-$3.03, $-$3.89, $-$4.59, and $-$8.21 eV, and Li-deficient and Li-excess environments. Panels (a)$-$(h) correspond to conditions (1)$-$(8), see text. Only the complex that consists of the lowest-energy negatively and positively charged point defects are included.}\label{fig;trends}
\end{figure*}

In order to help capture the most general trends in the energy landscape of native defects in LiFePO$_{4}$ in going from high to low $\mu_{{\rm O}_{2}}$ values and from Li-deficient to Li-excess environments, we plot in Fig.~\ref{fig;trends} the calculated formation energies of the most relevant native point defects and their lowest-energy complexes obtained under conditions (1)$-$(8). We find that, at a given Fermi-level position $\epsilon_{F}$, the formation energy of Fe$_{\rm{Li}}^+$ decreases as $\mu_{{\rm O}_{2}}$ decreases. This is because $\mu_{\rm Fe}$ increases more rapidly than $\mu_{\rm Li}$ does as $\mu_{{\rm O}_{2}}$ decreases from $-$3.03 eV to $-$8.21 eV. The formation energy of $p^{+}$, on the other hand, is independent of the choice of atomic chemical potentials, and depends only on $\epsilon_{F}$. At high $\mu_{{\rm O}_{2}}$ values, $p^{+}$ is lower in energy than Fe$_{\rm{Li}}^+$, but then the two defects switch orders before $\mu_{{\rm O}_{2}}$ reaches $-$3.89 eV (under Li-deficient environments) or $-$4.59 eV (Li-excess). The formation energy of the dominant positive defects, $p^{+}$ and Fe$_{\rm{Li}}^+$, differs by as much as 1.3 eV for some sets of atomic chemical potentials. On the contrary, the dominant negative defects, $V_{\rm{Li}}^-$ and Li$_{\rm{Fe}}^-$, have comparable formation energies throughout the range of conditions. The largest formation energy difference between the two defects is just 0.2 eV. The formation energy of Li$_{\rm{Fe}}^-$ is slightly lower than that of $V_{\rm{Li}}^-$ under Li-excess environments; whereas it is slightly higher under Li-deficient environments, except near $\mu_{{\rm O}_{2}}$=$-$8.21 eV where $V_{\rm{Li}}^-$ is higher than Li$_{\rm{Fe}}^-$. Both $V_{\rm{Li}}^-$ and Li$_{\rm{Fe}}^-$ have their formation energies increased as $\mu_{{\rm O}_{2}}$ decreases. In going from high to low $\mu_{{\rm O}_{2}}$ values, the changes in the formation energy of Fe$_{\rm{Li}}^+$ and that of $V_{\rm{Li}}^-$ and Li$_{\rm{Fe}}^-$ leads to a shift of the Fermi level from about 0.6 eV above the VBM to 2.0 eV above the VBM. The variation in the calculated formation energy of $p^{+}$ as seen in Table \ref{tab} is a result of this shift.

Under Li-deficient environments, we find that Fe$_{\rm{Li}}^+$ and $V_{\rm{Li}}^-$ are energetically most favorable over a wide range of $\mu_{{\rm O}_{2}}$ values, from $-$3.89 to $-$7.59 eV, and are likely to exist in form of the neutral complex Fe$_{\rm{Li}}^+$-$V_{\rm{Li}}^-$. At the higher end in the range of $\mu_{{\rm O}_{2}}$ values, from $-$3.03 to $-$3.25 eV, $p^{+}$ and $V_{\rm{Li}}^-$ are the most favorable, and are likely to exist in form of the complex $V_{\rm{Li}}^0$. Finally, only at the lowest end of the $\mu_{{\rm O}_{2}}$ values, the most favorable point defects are Fe$_{\rm{Li}}^+$ and Li$_{\rm{Fe}}^-$, which may exist in form of the neutral complex Fe$_{\rm{Li}}^+$-Li$_{\rm{Fe}}^-$. Under Li-excess environments, Li$_{\rm{Fe}}^-$ dominates the negatively charged point defects in the whole range of $\mu_{{\rm O}_{2}}$ values. This makes $p^{+}$ and Li$_{\rm{Fe}}^-$ the most favorable point defects for $\mu_{{\rm O}_{2}}$ ranging from $-$3.03 to $-$3.89 eV, and Fe$_{\rm{Li}}^+$ and Li$_{\rm{Fe}}^-$ the most favorable defects for $\mu_{{\rm O}_{2}}$ ranging from $-$4.59 to $-$8.21 eV. Note that, although the formation energy difference between $V_{\rm{Li}}^-$ and Li$_{\rm{Fe}}^-$ is small (less than 0.2 eV), the difference in their concentrations can still be significant, as indicated by the exponential dependence in Eq.~(\ref{eq;concen}).

Overall, we find that the calculated formation energies of the dominant native point defects are low, from about 0.3 to 0.5 eV for $\mu_{{\rm O}_{2}}$ from $-$3.03 to $-$8.21 eV ({\it cf.}~Table \ref{tab}). With such low formation energies, the defects will easily form and occur with high concentrations. The dominant defects may be different, however, if one changes the experimental conditions during synthesis, as discussed above. This is consistent with the reported experimental data showing the presence of various native defects in LiFePO$_{4}$ samples.\cite{Ellis2006,Zaghib2007,Yang2002,maier2008,chen2008,Chung2008,Axmann:2009p137,Chung2010} We note that there are several limitations inherent in our calculations. The first set of limitations comes from standard methodological uncertainties contained in the calculated formation enthalpies and phase diagrams as discussed in Ref.\cite{ong2008}. The second set comes from the calculation of defect formation energies using supercell models where supercell finite-size effects are expected.\cite{walle:3851} Since applying approximations indiscriminately in an attempt to correct for finite-size effects tends to ``overshoot'' and makes the energies even less accurate,\cite{walle:3851} we did not include any corrections pertaining to such effects in our defect calculations. A proper treatment of finite-size effects, however if applied, will lead to an increase in the calculated formation energy of the charged native point defects, and hence the binding energy of the neutral defect complexes. In spite of the limitations, the general trends discussed above should still hold true. And, our results therefore can serve as guidelines for tailoring the defect concentrations in LiFePO$_{4}$, and suppressing or enhancing certain defects for targeted applications. 

\section{Electronic and Ionic Conduction}

Strictly speaking, lithium vacancies are only stable as $V_{\rm{Li}}^-$, and $V_{\rm{Li}}^0$ (which is, in fact, a complex of $V_{\rm{Li}}^-$ and $p^{+}$) cannot be considered the vacancies's neutral charge state. Likewise, lithium antisites, iron antisites, and iron vacancies also have one stable charge state only and occur as, respectively, Li$_{\rm{Fe}}^{-}$, Fe$_{\rm{Li}}^{+}$, and $V_{\rm{Fe}}^{2-}$. Removing/adding electrons from/to these stable point defects always results in defect complexes consisting of the point defects and small hole/electron polarons, as presented in the previous sections. The fact that small polarons can be stabilized, both in the absence and in the presence of other native defects is necessarily related to the electronic structure of LiFePO$_{4}$ where the VBM and CBM consist predominantly of the highly localized Fe 3$d$ states. Combined with the fact that charged native defects have negative formation energies near the VBM and CBM ({\it cf.}~Figures 3, 5, and 6), our results therefore indicate that native defects in LiFePO$_{4}$ cannot act as sources of band-like hole and electron conductivities. These defects will, however, act as compensating centers in donor-like doping (for the negatively charged defects) or acceptor-like doping (the positively charged defects). The electronic conduction in LiFePO$_{4}$ thus occurs via hopping of small hole polarons. This mechanism, in fact, has been proposed for LiFePO$_{4}$ in several previous works.\cite{Zhou:2004p101,Maxisch:2006p103,Ellis2006,Zaghib2007} Zaghib et al.\cite{Zaghib2007} found experimental evidence of intra-atomic Fe$^{2+}$$-$Fe$^{3+}$ transitions in the optical spectrum of LiFePO$_{4}$, which indirectly confirms the formation of small hole polarons. The activation energy for electronic conductivity in LiFePO$_{4}$ was estimated to be 0.65 eV,\cite{Zaghib2007} comparable to that of 0.55$-$0.78 eV reported by several other experimental groups.\cite{delacourt:A913,Ellis2006,Amin2008} 

In order to compare our results with the measured activation energies, let us assume two scenarios for hole polaron hopping in LiFePO$_{4}$. In the first scenario, we assume self-diffusion of free $p^{+}$ defects. The activation energy $E_{a}$ for this process is calculated as the summation of the formation energy and migration barrier of $p^{+}$, where the former is associated with the intrinsic concentration and the latter with the mobility,\cite{balluffi2005kinetics} 
\begin{equation}\label{eq;Ea1} 
E_{a} = E^{f}(p^{+}) + E_{m}(p^{+}), 
\end{equation} 
which gives $E_{a}$=0.57 eV, if $E^{f}(p^{+})$ is taken under the most favorable condition where $p^{+}$ has the lowest formation energy ({\it cf.}~Table \ref{tab}). In the second scenario, we assume that $p^{+}$ and $V_{\rm{Li}}^{-}$ are formed via the formation of the neutral complex $V_{\rm{Li}}^{0}$, similar to cases where defects are created via a Frenkel or Schottky mechanism.\cite{balluffi2005kinetics} At high temperatures, the activation energy for the diffusion of $p^{+}$ can be calculated as 
\begin{equation}\label{eq;Ea2} 
E_{a} = \frac{1}{2}E^{f}(V_{\rm{Li}}^{0}) + E_{m}(p^{+}), 
\end{equation} 
which results in $E_{a}$=0.41 eV, assuming the condition where $V_{\rm{Li}}^{0}$ has the lowest formation energy ({\it cf.}~Table \ref{tab}). The lower bound of the activation energy for polaron conductivity is therefore 0.41$-$0.57 eV. This range of $E_{a}$ values is comparable to that obtained in experiments.\cite{delacourt:A913,Ellis2006,Zaghib2007,Amin2008}

Among the native defects, $V_{\rm{Li}}^{-}$ is the most plausible candidate for ionic conduction in LiFePO$_{4}$, because of its low formation energy and high mobility. Using formulae similar to Eqs.~(7) and (8), we estimate the lower bound of the activation energy for self-diffusion of $V_{\rm{Li}}^{-}$ along the $b$-axis to be 0.48$-$0.65 eV. The reported experimental value is 0.54 eV\cite{Li2008} or 0.62$-$0.74 eV,\cite{Amin2008} obtained from ionic conductivity measurements carried out on LiFePO$_{4}$ single crystals, which is comparable to our calculated value. The diffusion of $V_{\rm{Li}}^{-}$, however, may be impeded by other native defects or extrinsic impurities that have lower mobility. The presence of Fe$_{\rm Li}^{+}$ in LiFePO$_{4}$ has been reported and the defect is believed to reduce the electrochemical performance of the material by blocking the lithium channels.\cite{Yang2002,maier2008,chen2008,Chung2008,Axmann:2009p137,Chung2010} Indeed, our results also show that Fe$_{\rm Li}^{+}$ occurs with a high concentration under various conditions ({\it cf.}~Table \ref{tab}) and has low mobility. Whether Fe$_{\rm Li}^{+}$ is stable in form of the neutral complex Fe$_{\rm{Li}}^+$-$V_{\rm{Li}}^-$ or Fe$_{\rm{Li}}^+$-Li$_{\rm{Fe}}^-$ as suggested in several experimental works,\cite{maier2008,Chung2008,Axmann:2009p137,Chung2010} however, depends on the specific conditions under which the samples are produced.

What is most fascinating about our results is that one can suppress Fe$_{\rm{Li}}^+$ by adjusting suitable experimental conditions during synthesis. In fact, $p^{+}$-rich and Fe$_{\rm Li}^{+}$-free LiFePO$_{4}$ samples, which are believed to be desirable for high intrinsic electronic and ionic conductivities, can be produced if one maintains high $\mu_{{\rm O}_{2}}$ values ({\it cf.}~Fig.~\ref{fig;trends}). Of course, $\mu_{{\rm O}_{2}}$ should not be so high that LiFePO$_{4}$ becomes unstable toward forming secondary phases. Although LiFePO$_{4}$ cannot be doped $p$-type or $n$-type as discussed earlier, the incorporation of suitable electrically active impurities in the material can enhance the electronic (ionic) conductivity via increasing the concentration of $p^{+}$ ($V_{\rm{Li}}^{-}$). These impurities, {\it if} present in the samples with a concentration higher than that of the charged native defects, can shift the Fermi level,\cite{peles2007,hoang2009,wilson-short} and hence lower the formation energy of either $p^{+}$ or $V_{\rm{Li}}^{-}$. A decrease in the formation energy of $p^{+}$, however, may result in an increase in the formation energy of $V_{\rm{Li}}^{-}$ and vice versa. For example, impurities with positive effective charges (i.e., donor-like doping) may shift the Fermi level to the right ({\it cf.}~Fig.~\ref{fig;trends}), resulting in an increased (decreased) formation energy of $p^{+}$ ($V_{\rm{Li}}^{-}$). Impurities with negative effective charges (i.e., acceptor-like doping), on the other hand, may produce the opposite effects, namely, decreasing (increasing) the formation energy of $p^{+}$ ($V_{\rm{Li}}^{-}$). An enhancement in both electronic and ionic conductivities would, therefore, require a delicate combination of defect-controlled synthesis, doping with suitable electrically active impurities, and post-synthesis treatments. An example of the latter would be thermal treatment which, in fact, has been reported to cause lithium loss in LiFePO$_{4}$ and lower the activation energy of the electrical conductivity.\cite{Amin20081831}

\section{\label{sec;conclusion}Conclusion}

In summary, we have carried out comprehensive first-principles studies of native point defects and defect complexes in LiFePO$_{4}$. We find that lithium vacancies, lithium antisites, iron antisites, and iron vacancies each have one stable charge state only and occur as, respectively, $V_{\rm{Li}}^{-}$, Li$_{\rm{Fe}}^{-}$, Fe$_{\rm{Li}}^{+}$, and $V_{\rm{Fe}}^{2-}$. The removal/addition of electrons from/to these stable native point defects does not result in a transition to other charge states of the same defects, but instead generates small hole/electron polarons. The fact that small polarons can be stabilized, both in the presence and in the absence of other native defects, is necessarily related to the electronic structure of LiFePO$_{4}$. Our analysis thus indicates that native defects in the material cannot act as sources of band-like electron and hole conductivities, and the electronic conduction, in fact, proceeds via hopping of small hole polarons ($p^{+}$). The ionic conduction, on the other hand, occurs via diffusion of lithium vacancies.

Among all possible native defects, $p^{+}$, $V_{\rm{Li}}^{-}$, Li$_{\rm{Fe}}^{-}$, and Fe$_{\rm{Li}}^{+}$ are found to have low formation energies and are hence expected to occur in LiFePO$_{4}$ with high concentrations. The dominant point defects in the samples are likely to exist in forms of a neutral defect complex such as $V_{\rm{Li}}^{0}$, Li$_{\rm{Fe}}^{0}$, Fe$_{\rm{Li}}^{+}$-$V_{\rm{Li}}^{-}$, or Fe$_{\rm{Li}}^{+}$-Li$_{\rm{Fe}}^{-}$. The energy landscape of these defects is, however, sensitive to the choice of atomic chemical potentials which represent experimental conditions during synthesis. This explains the conflicting experimental data on defect formation in LiFePO$_{4}$. Our results also raise the necessity of having prior knowledge of the native defects in LiFePO$_{4}$ samples before any useful interpretations of the measured transport data can be made. We suggest that one can suppress or enhance certain native defects in LiFePO$_{4}$ via tuning the experimental conditions during synthesis, and thereby produce samples with tailored defect concentrations for optimal performance. The electrical conductivity may be enhanced through increase of hole polaron and lithium vacancy concentrations via a combination of defect-controlled synthesis, incorporation of suitable electrically active impurities that can shift the Fermi level, and post-synthesis treatments.

\begin{acknowledgments} 

The authors acknowledge helpful discussions with S. C. Erwin and J. Allen, and the use of computing facilities at the DoD HPC Centers. K.H.~was supported by Naval Research Laboratory through Grant No.~NRL-N00173-08-G001, and M.J.~by the Office of Naval Research.

\end{acknowledgments}

\begin{mcitethebibliography}{55}
\providecommand*\natexlab[1]{#1}
\providecommand*\mciteSetBstSublistMode[1]{}
\providecommand*\mciteSetBstMaxWidthForm[2]{}
\providecommand*\mciteBstWouldAddEndPuncttrue
  {\def\EndOfBibitem{\unskip.}}
\providecommand*\mciteBstWouldAddEndPunctfalse
  {\let\EndOfBibitem\relax}
\providecommand*\mciteSetBstMidEndSepPunct[3]{}
\providecommand*\mciteSetBstSublistLabelBeginEnd[3]{}
\providecommand*\EndOfBibitem{}
\mciteSetBstSublistMode{f}
\mciteSetBstMaxWidthForm{subitem}{(\alph{mcitesubitemcount})}
\mciteSetBstSublistLabelBeginEnd
  {\mcitemaxwidthsubitemform\space}
  {\relax}
  {\relax}

\bibitem[Padhi et~al.(1997)Padhi, Nanjundaswamy, and Goodenough]{padhi:1188}
Padhi,~A.~K.; Nanjundaswamy,~K.~S.; Goodenough,~J.~B. \emph{J. Electrochem.
  Soc.} \textbf{1997}, \emph{144}, 1188--1194\relax
\mciteBstWouldAddEndPuncttrue
\mciteSetBstMidEndSepPunct{\mcitedefaultmidpunct}
{\mcitedefaultendpunct}{\mcitedefaultseppunct}\relax
\EndOfBibitem
\bibitem[Ellis et~al.(2010)Ellis, Lee, and Nazar]{Ellis2010}
Ellis,~B.~L.; Lee,~K.~T.; Nazar,~L.~F. \emph{Chem. Mater.} \textbf{2010},
  \emph{22}, 691--714\relax
\mciteBstWouldAddEndPuncttrue
\mciteSetBstMidEndSepPunct{\mcitedefaultmidpunct}
{\mcitedefaultendpunct}{\mcitedefaultseppunct}\relax
\EndOfBibitem
\bibitem[Manthiram(2011)]{Manthiram2011}
Manthiram,~A. \emph{J. Phys. Chem. Lett.} \textbf{2011}, \emph{2},
  176--184\relax
\mciteBstWouldAddEndPuncttrue
\mciteSetBstMidEndSepPunct{\mcitedefaultmidpunct}
{\mcitedefaultendpunct}{\mcitedefaultseppunct}\relax
\EndOfBibitem
\bibitem[Delacourt et~al.(2005)Delacourt, Laffont, Bouchet, Wurm, Leriche,
  Morcrette, Tarascon, and Masquelier]{delacourt:A913}
Delacourt,~C.; Laffont,~L.; Bouchet,~R.; Wurm,~C.; Leriche,~J.-B.;
  Morcrette,~M.; Tarascon,~J.-M.; Masquelier,~C. \emph{J. Electrochem. Soc.}
  \textbf{2005}, \emph{152}, A913--A921\relax
\mciteBstWouldAddEndPuncttrue
\mciteSetBstMidEndSepPunct{\mcitedefaultmidpunct}
{\mcitedefaultendpunct}{\mcitedefaultseppunct}\relax
\EndOfBibitem
\bibitem[Huang et~al.(2001)Huang, Yin, and Nazar]{huang:A170}
Huang,~H.; Yin,~S.-C.; Nazar,~L.~F. \emph{Electrochem. Solid-State Lett.}
  \textbf{2001}, \emph{4}, A170--A172\relax
\mciteBstWouldAddEndPuncttrue
\mciteSetBstMidEndSepPunct{\mcitedefaultmidpunct}
{\mcitedefaultendpunct}{\mcitedefaultseppunct}\relax
\EndOfBibitem
\bibitem[Ellis et~al.(2007)Ellis, Kan, Makahnouk, and Nazar]{Ellis2007}
Ellis,~B.; Kan,~W.~H.; Makahnouk,~W. R.~M.; Nazar,~L.~F. \emph{J. Mater. Chem.}
  \textbf{2007}, \emph{17}, 3248--3254\relax
\mciteBstWouldAddEndPuncttrue
\mciteSetBstMidEndSepPunct{\mcitedefaultmidpunct}
{\mcitedefaultendpunct}{\mcitedefaultseppunct}\relax
\EndOfBibitem
\bibitem[Chung et~al.(2002)Chung, Bloking, and Chiang]{Chung:2002p246}
Chung,~S.; Bloking,~J.; Chiang,~Y. \emph{Nat. Mater.} \textbf{2002}, \emph{1},
  123--128\relax
\mciteBstWouldAddEndPuncttrue
\mciteSetBstMidEndSepPunct{\mcitedefaultmidpunct}
{\mcitedefaultendpunct}{\mcitedefaultseppunct}\relax
\EndOfBibitem
\bibitem[Ravet et~al.({2003})Ravet, Abouimrane, and Armand]{Ravet2003}
Ravet,~N.; Abouimrane,~A.; Armand,~M. \emph{{Nat. Mater.}} \textbf{{2003}},
  \emph{{2}}, {702}\relax
\mciteBstWouldAddEndPuncttrue
\mciteSetBstMidEndSepPunct{\mcitedefaultmidpunct}
{\mcitedefaultendpunct}{\mcitedefaultseppunct}\relax
\EndOfBibitem
\bibitem[Wagemaker et~al.(2008)Wagemaker, Ellis, L\"{u}tzenkirchen-Hecht,
  Mulder, and Nazar]{Wagemaker2008}
Wagemaker,~M.; Ellis,~B.~L.; L\"{u}tzenkirchen-Hecht,~D.; Mulder,~F.~M.;
  Nazar,~L.~F. \emph{Chem. Mater.} \textbf{2008}, \emph{20}, 6313--6315\relax
\mciteBstWouldAddEndPuncttrue
\mciteSetBstMidEndSepPunct{\mcitedefaultmidpunct}
{\mcitedefaultendpunct}{\mcitedefaultseppunct}\relax
\EndOfBibitem
\bibitem[Zhou et~al.(2004)Zhou, Kang, Maxisch, Ceder, and
  Morgan]{Zhou:2004p101}
Zhou,~F.; Kang,~K.; Maxisch,~T.; Ceder,~G.; Morgan,~D. \emph{Solid State
  Commun.} \textbf{2004}, \emph{132}, 181--186\relax
\mciteBstWouldAddEndPuncttrue
\mciteSetBstMidEndSepPunct{\mcitedefaultmidpunct}
{\mcitedefaultendpunct}{\mcitedefaultseppunct}\relax
\EndOfBibitem
\bibitem[Maxisch et~al.(2006)Maxisch, Zhou, and Ceder]{Maxisch:2006p103}
Maxisch,~T.; Zhou,~F.; Ceder,~G. \emph{Phys. Rev. B} \textbf{2006}, \emph{73},
  104301\relax
\mciteBstWouldAddEndPuncttrue
\mciteSetBstMidEndSepPunct{\mcitedefaultmidpunct}
{\mcitedefaultendpunct}{\mcitedefaultseppunct}\relax
\EndOfBibitem
\bibitem[Ellis et~al.(2006)Ellis, Perry, Ryan, and Nazar]{Ellis2006}
Ellis,~B.; Perry,~L.~K.; Ryan,~D.~H.; Nazar,~L.~F. \emph{J. Am. Chem. Soc.}
  \textbf{2006}, \emph{128}, 11416--11422\relax
\mciteBstWouldAddEndPuncttrue
\mciteSetBstMidEndSepPunct{\mcitedefaultmidpunct}
{\mcitedefaultendpunct}{\mcitedefaultseppunct}\relax
\EndOfBibitem
\bibitem[Zaghib et~al.(2007)Zaghib, Mauger, Goodenough, Gendron, and
  Julien]{Zaghib2007}
Zaghib,~K.; Mauger,~A.; Goodenough,~J.~B.; Gendron,~F.; Julien,~C.~M.
  \emph{Chem. Mater.} \textbf{2007}, \emph{19}, 3740--3747\relax
\mciteBstWouldAddEndPuncttrue
\mciteSetBstMidEndSepPunct{\mcitedefaultmidpunct}
{\mcitedefaultendpunct}{\mcitedefaultseppunct}\relax
\EndOfBibitem
\bibitem[Yang et~al.({2002})Yang, Song, Zavalij, and Whittingham]{Yang2002}
Yang,~S.~F.; Song,~Y.~N.; Zavalij,~P.~Y.; Whittingham,~M.~S.
  \emph{{Electrochem. Commun.}} \textbf{{2002}}, \emph{{4}}, {239--244}\relax
\mciteBstWouldAddEndPuncttrue
\mciteSetBstMidEndSepPunct{\mcitedefaultmidpunct}
{\mcitedefaultendpunct}{\mcitedefaultseppunct}\relax
\EndOfBibitem
\bibitem[Maier and Amin(2008)Maier, and Amin]{maier2008}
Maier,~J.; Amin,~R. \emph{J. Electrochem. Soc.} \textbf{2008}, \emph{155},
  A339--A344\relax
\mciteBstWouldAddEndPuncttrue
\mciteSetBstMidEndSepPunct{\mcitedefaultmidpunct}
{\mcitedefaultendpunct}{\mcitedefaultseppunct}\relax
\EndOfBibitem
\bibitem[Chen et~al.(2008)Chen, Vacchio, Wang, Chernova, Zavalij, and
  Whittingham]{chen2008}
Chen,~J.~J.; Vacchio,~M.~J.; Wang,~S.~J.; Chernova,~N.; Zavalij,~P.~Y.;
  Whittingham,~M.~S. \emph{Solid State Ionics} \textbf{2008}, \emph{178},
  1676--1693\relax
\mciteBstWouldAddEndPuncttrue
\mciteSetBstMidEndSepPunct{\mcitedefaultmidpunct}
{\mcitedefaultendpunct}{\mcitedefaultseppunct}\relax
\EndOfBibitem
\bibitem[Chung et~al.(2008)Chung, Choi, Yamamoto, and Ikuhara]{Chung2008}
Chung,~S.-Y.; Choi,~S.-Y.; Yamamoto,~T.; Ikuhara,~Y. \emph{Phys. Rev. Lett.}
  \textbf{2008}, \emph{100}, 125502\relax
\mciteBstWouldAddEndPuncttrue
\mciteSetBstMidEndSepPunct{\mcitedefaultmidpunct}
{\mcitedefaultendpunct}{\mcitedefaultseppunct}\relax
\EndOfBibitem
\bibitem[Axmann et~al.(2009)Axmann, Stinner, Wohlfahrt-Mehrens, Mauger,
  Gendron, and Julien]{Axmann:2009p137}
Axmann,~P.; Stinner,~C.; Wohlfahrt-Mehrens,~M.; Mauger,~A.; Gendron,~F.;
  Julien,~C.~M. \emph{Chem. Mater.} \textbf{2009}, \emph{21}, 1636--1644\relax
\mciteBstWouldAddEndPuncttrue
\mciteSetBstMidEndSepPunct{\mcitedefaultmidpunct}
{\mcitedefaultendpunct}{\mcitedefaultseppunct}\relax
\EndOfBibitem
\bibitem[Chung et~al.(2010)Chung, Kim, and Choi]{Chung2010}
Chung,~S.-Y.; Kim,~Y.-M.; Choi,~S.-Y. \emph{Adv. Funct. Mater.} \textbf{2010},
  \emph{20}, 4219--4232\relax
\mciteBstWouldAddEndPuncttrue
\mciteSetBstMidEndSepPunct{\mcitedefaultmidpunct}
{\mcitedefaultendpunct}{\mcitedefaultseppunct}\relax
\EndOfBibitem
\bibitem[Morgan et~al.(2004)Morgan, {Van der Ven}, and Ceder]{morgan:A30}
Morgan,~D.; {Van der Ven},~A.; Ceder,~G. \emph{Electrochem. Solid-State Lett.}
  \textbf{2004}, \emph{7}, A30--A32\relax
\mciteBstWouldAddEndPuncttrue
\mciteSetBstMidEndSepPunct{\mcitedefaultmidpunct}
{\mcitedefaultendpunct}{\mcitedefaultseppunct}\relax
\EndOfBibitem
\bibitem[Islam et~al.(2005)Islam, Driscoll, Fisher, and Slater]{Islam:2005p168}
Islam,~M.~S.; Driscoll,~D.~J.; Fisher,~C. A.~J.; Slater,~P.~R. \emph{Chem.
  Mater.} \textbf{2005}, \emph{17}, 5085--5092\relax
\mciteBstWouldAddEndPuncttrue
\mciteSetBstMidEndSepPunct{\mcitedefaultmidpunct}
{\mcitedefaultendpunct}{\mcitedefaultseppunct}\relax
\EndOfBibitem
\bibitem[Fisher et~al.(2008)Fisher, Prieto, and Islam]{Fisher:2008p80}
Fisher,~C. A.~J.; Prieto,~V. M.~H.; Islam,~M.~S. \emph{Chem. Mater.}
  \textbf{2008}, \emph{20}, 5907--5915\relax
\mciteBstWouldAddEndPuncttrue
\mciteSetBstMidEndSepPunct{\mcitedefaultmidpunct}
{\mcitedefaultendpunct}{\mcitedefaultseppunct}\relax
\EndOfBibitem
\bibitem[Adams(2010)]{Adams:2010p132}
Adams,~S. \emph{J. Solid State Electrochem.} \textbf{2010}, \emph{14},
  1787--1792\relax
\mciteBstWouldAddEndPuncttrue
\mciteSetBstMidEndSepPunct{\mcitedefaultmidpunct}
{\mcitedefaultendpunct}{\mcitedefaultseppunct}\relax
\EndOfBibitem
\bibitem[Malik et~al.(2010)Malik, Burch, Bazant, and Ceder]{Malik2010}
Malik,~R.; Burch,~D.; Bazant,~M.; Ceder,~G. \emph{Nano Letters} \textbf{2010},
  \emph{10}, 4123--4127\relax
\mciteBstWouldAddEndPuncttrue
\mciteSetBstMidEndSepPunct{\mcitedefaultmidpunct}
{\mcitedefaultendpunct}{\mcitedefaultseppunct}\relax
\EndOfBibitem
\bibitem[Anisimov et~al.(1991)Anisimov, Zaanen, and Andersen]{anisimov1991}
Anisimov,~V.~I.; Zaanen,~J.; Andersen,~O.~K. \emph{Phys. Rev. B} \textbf{1991},
  \emph{44}, 943--954\relax
\mciteBstWouldAddEndPuncttrue
\mciteSetBstMidEndSepPunct{\mcitedefaultmidpunct}
{\mcitedefaultendpunct}{\mcitedefaultseppunct}\relax
\EndOfBibitem
\bibitem[Anisimov et~al.(1993)Anisimov, Solovyev, Korotin,
  Czy\ifmmode~\dot{z}\else \.{z}\fi{}yk, and Sawatzky]{anisimov1993}
Anisimov,~V.~I.; Solovyev,~I.~V.; Korotin,~M.~A.; Czy\ifmmode~\dot{z}\else
  \.{z}\fi{}yk,~M.~T.; Sawatzky,~G.~A. \emph{Phys. Rev. B} \textbf{1993},
  \emph{48}, 16929--16934\relax
\mciteBstWouldAddEndPuncttrue
\mciteSetBstMidEndSepPunct{\mcitedefaultmidpunct}
{\mcitedefaultendpunct}{\mcitedefaultseppunct}\relax
\EndOfBibitem
\bibitem[Liechtenstein et~al.(1995)Liechtenstein, Anisimov, and
  Zaanen]{liechtenstein1995}
Liechtenstein,~A.~I.; Anisimov,~V.~I.; Zaanen,~J. \emph{Phys. Rev. B}
  \textbf{1995}, \emph{52}, R5467--R5470\relax
\mciteBstWouldAddEndPuncttrue
\mciteSetBstMidEndSepPunct{\mcitedefaultmidpunct}
{\mcitedefaultendpunct}{\mcitedefaultseppunct}\relax
\EndOfBibitem
\bibitem[Perdew et~al.(1996)Perdew, Burke, and Ernzerhof]{GGA}
Perdew,~J.~P.; Burke,~K.; Ernzerhof,~M. \emph{Phys. Rev. Lett.} \textbf{1996},
  \emph{77}, 3865--3868\relax
\mciteBstWouldAddEndPuncttrue
\mciteSetBstMidEndSepPunct{\mcitedefaultmidpunct}
{\mcitedefaultendpunct}{\mcitedefaultseppunct}\relax
\EndOfBibitem
\bibitem[Bl\"ochl(1994)]{PAW1}
Bl\"ochl,~P.~E. \emph{Phys. Rev. B} \textbf{1994}, \emph{50},
  17953--17979\relax
\mciteBstWouldAddEndPuncttrue
\mciteSetBstMidEndSepPunct{\mcitedefaultmidpunct}
{\mcitedefaultendpunct}{\mcitedefaultseppunct}\relax
\EndOfBibitem
\bibitem[Kresse and Joubert(1999)Kresse, and Joubert]{PAW2}
Kresse,~G.; Joubert,~D. \emph{Phys. Rev. B} \textbf{1999}, \emph{59},
  1758--1775\relax
\mciteBstWouldAddEndPuncttrue
\mciteSetBstMidEndSepPunct{\mcitedefaultmidpunct}
{\mcitedefaultendpunct}{\mcitedefaultseppunct}\relax
\EndOfBibitem
\bibitem[Kresse and Hafner(1993)Kresse, and Hafner]{VASP1}
Kresse,~G.; Hafner,~J. \emph{Phys. Rev. B} \textbf{1993}, \emph{47},
  558--561\relax
\mciteBstWouldAddEndPuncttrue
\mciteSetBstMidEndSepPunct{\mcitedefaultmidpunct}
{\mcitedefaultendpunct}{\mcitedefaultseppunct}\relax
\EndOfBibitem
\bibitem[Kresse and Furthm\"uller(1996)Kresse, and Furthm\"uller]{VASP2}
Kresse,~G.; Furthm\"uller,~J. \emph{Phys. Rev. B} \textbf{1996}, \emph{54},
  11169--11186\relax
\mciteBstWouldAddEndPuncttrue
\mciteSetBstMidEndSepPunct{\mcitedefaultmidpunct}
{\mcitedefaultendpunct}{\mcitedefaultseppunct}\relax
\EndOfBibitem
\bibitem[Kresse and Furthmüller(1996)Kresse, and Furthmüller]{VASP3}
Kresse,~G.; Furthmüller,~J. \emph{Comput. Mat. Sci.} \textbf{1996}, \emph{6},
  15--50\relax
\mciteBstWouldAddEndPuncttrue
\mciteSetBstMidEndSepPunct{\mcitedefaultmidpunct}
{\mcitedefaultendpunct}{\mcitedefaultseppunct}\relax
\EndOfBibitem
\bibitem[Zhou et~al.(2004)Zhou, Cococcioni, Marianetti, Morgan, and
  Ceder]{Zhou:2004p104}
Zhou,~F.; Cococcioni,~M.; Marianetti,~C.; Morgan,~D.; Ceder,~G. \emph{Phys.
  Rev. B} \textbf{2004}, \emph{70}, 235121\relax
\mciteBstWouldAddEndPuncttrue
\mciteSetBstMidEndSepPunct{\mcitedefaultmidpunct}
{\mcitedefaultendpunct}{\mcitedefaultseppunct}\relax
\EndOfBibitem
\bibitem[Monkhorst and Pack(1976)Monkhorst, and Pack]{monkhorst-pack}
Monkhorst,~H.~J.; Pack,~J.~D. \emph{Phys. Rev. B} \textbf{1976}, \emph{13},
  5188--5192\relax
\mciteBstWouldAddEndPuncttrue
\mciteSetBstMidEndSepPunct{\mcitedefaultmidpunct}
{\mcitedefaultendpunct}{\mcitedefaultseppunct}\relax
\EndOfBibitem
\bibitem[Henkelman et~al.(2000)Henkelman, Uberuaga, and J\'{o}nsson]{ci-neb}
Henkelman,~G.; Uberuaga,~B.~P.; J\'{o}nsson,~H. \emph{J. Chem. Phys.}
  \textbf{2000}, \emph{113}, 9901--9904\relax
\mciteBstWouldAddEndPuncttrue
\mciteSetBstMidEndSepPunct{\mcitedefaultmidpunct}
{\mcitedefaultendpunct}{\mcitedefaultseppunct}\relax
\EndOfBibitem
\bibitem[Rousse et~al.(2003)Rousse, Rodriguez-Carvajal, Patoux, and
  Masquelier]{rousse2003}
Rousse,~G.; Rodriguez-Carvajal,~J.; Patoux,~S.; Masquelier,~C. \emph{Chem.
  Mater.} \textbf{2003}, \emph{15}, 4082--4090\relax
\mciteBstWouldAddEndPuncttrue
\mciteSetBstMidEndSepPunct{\mcitedefaultmidpunct}
{\mcitedefaultendpunct}{\mcitedefaultseppunct}\relax
\EndOfBibitem
\bibitem[{Van de Walle} and Neugebauer(2004){Van de Walle}, and
  Neugebauer]{walle:3851}
{Van de Walle},~C.~G.; Neugebauer,~J. \emph{J. Appl. Phys.} \textbf{2004},
  \emph{95}, 3851--3879\relax
\mciteBstWouldAddEndPuncttrue
\mciteSetBstMidEndSepPunct{\mcitedefaultmidpunct}
{\mcitedefaultendpunct}{\mcitedefaultseppunct}\relax
\EndOfBibitem
\bibitem[Janotti and {Van de Walle}(2009)Janotti, and {Van de
  Walle}]{janotti2009}
Janotti,~A.; {Van de Walle},~C.~G. \emph{Rep. Prog. Phys.} \textbf{2009},
  \emph{72}, 126501\relax
\mciteBstWouldAddEndPuncttrue
\mciteSetBstMidEndSepPunct{\mcitedefaultmidpunct}
{\mcitedefaultendpunct}{\mcitedefaultseppunct}\relax
\EndOfBibitem
\bibitem[{Van de Walle} and Janotti(2010){Van de Walle}, and Janotti]{vdW2010}
{Van de Walle},~C.~G.; Janotti,~A. \emph{IOP Conf. Ser.: Mater. Sci. Eng.}
  \textbf{2010}, \emph{15}, 012001\relax
\mciteBstWouldAddEndPuncttrue
\mciteSetBstMidEndSepPunct{\mcitedefaultmidpunct}
{\mcitedefaultendpunct}{\mcitedefaultseppunct}\relax
\EndOfBibitem
\bibitem[Ong et~al.(2008)Ong, Wang, Kang, and Ceder]{ong2008}
Ong,~P.~S.; Wang,~L.; Kang,~B.; Ceder,~G. \emph{Chem. Mater.} \textbf{2008},
  \emph{20}, 1798--1807\relax
\mciteBstWouldAddEndPuncttrue
\mciteSetBstMidEndSepPunct{\mcitedefaultmidpunct}
{\mcitedefaultendpunct}{\mcitedefaultseppunct}\relax
\EndOfBibitem
\bibitem[Not()]{Note-1}
Note that we have applied a shift of 1.36 eV per O$_{2}$ molecule to $E_{{\rm
  O}_{2}}^{\rm tot}$ as suggested by Wang et al.~[Wang, L.; Maxisch, T.; Ceder,
  G. {\it Phys. Rev. B} {\bf 2006}, {\it 73}, 195107] and discussed in
  Ref.\cite{ong2008}. This constant shift is to correct for the O$_{2}$ binding
  energy error and the error in charge transfer ($d$$\rightarrow$$p$) energy
  due to improper treatment of correlation.\relax
\mciteBstWouldAddEndPunctfalse
\mciteSetBstMidEndSepPunct{\mcitedefaultmidpunct}
{}{\mcitedefaultseppunct}\relax
\EndOfBibitem
\bibitem[Peles and {Van de Walle}(2007)Peles, and {Van de Walle}]{peles2007}
Peles,~A.; {Van de Walle},~C.~G. \emph{Phys. Rev. B} \textbf{2007}, \emph{76},
  214101\relax
\mciteBstWouldAddEndPuncttrue
\mciteSetBstMidEndSepPunct{\mcitedefaultmidpunct}
{\mcitedefaultendpunct}{\mcitedefaultseppunct}\relax
\EndOfBibitem
\bibitem[Hoang and {Van de Walle}(2009)Hoang, and {Van de Walle}]{hoang2009}
Hoang,~K.; {Van de Walle},~C.~G. \emph{Phys. Rev. B} \textbf{2009}, \emph{80},
  214109\relax
\mciteBstWouldAddEndPuncttrue
\mciteSetBstMidEndSepPunct{\mcitedefaultmidpunct}
{\mcitedefaultendpunct}{\mcitedefaultseppunct}\relax
\EndOfBibitem
\bibitem[Wilson-Short et~al.(2009)Wilson-Short, Janotti, Hoang, Peles, and {Van
  de Walle}]{wilson-short}
Wilson-Short,~G.~B.; Janotti,~A.; Hoang,~K.; Peles,~A.; {Van de Walle},~C.~G.
  \emph{Phys. Rev. B} \textbf{2009}, \emph{80}, 224102\relax
\mciteBstWouldAddEndPuncttrue
\mciteSetBstMidEndSepPunct{\mcitedefaultmidpunct}
{\mcitedefaultendpunct}{\mcitedefaultseppunct}\relax
\EndOfBibitem
\bibitem[{Van de Walle} et~al.(2010){Van de Walle}, Lyons, and
  Janotti]{vdwalle2010}
{Van de Walle},~C.~G.; Lyons,~J.~L.; Janotti,~A. \emph{phys. status solidi (a)}
  \textbf{2010}, \emph{207}, 1024--1036\relax
\mciteBstWouldAddEndPuncttrue
\mciteSetBstMidEndSepPunct{\mcitedefaultmidpunct}
{\mcitedefaultendpunct}{\mcitedefaultseppunct}\relax
\EndOfBibitem
\bibitem[{Catlow, C. R. A.; Sokol, A. A.; Walsh, A.}({2011})]{catlow2011}
{Catlow, C. R. A.; Sokol, A. A.; Walsh, A.}, \emph{{Chem. Commun.}}
  \textbf{{2011}}, \emph{{47}}, {3386--3388}\relax
\mciteBstWouldAddEndPuncttrue
\mciteSetBstMidEndSepPunct{\mcitedefaultmidpunct}
{\mcitedefaultendpunct}{\mcitedefaultseppunct}\relax
\EndOfBibitem
\bibitem[Shluger and Stoneham(1993)Shluger, and Stoneham]{Shluger1993}
Shluger,~A.~L.; Stoneham,~A.~M. \emph{J. Phys.: Condens. Matter} \textbf{1993},
  \emph{5}, 3049--3086\relax
\mciteBstWouldAddEndPuncttrue
\mciteSetBstMidEndSepPunct{\mcitedefaultmidpunct}
{\mcitedefaultendpunct}{\mcitedefaultseppunct}\relax
\EndOfBibitem
\bibitem[Stoneham et~al.(2007)Stoneham, Gavartin, Shluger, Kimmel, {Mu{\~{n}}oz
  Ramo}, R{\o}nnow, Aeppli, and Renner]{Stoneham2007}
Stoneham,~A.~M.; Gavartin,~J.; Shluger,~A.~L.; Kimmel,~A.~V.; {Mu{\~{n}}oz
  Ramo},~D.; R{\o}nnow,~H.~M.; Aeppli,~G.; Renner,~C. \emph{J. Phys.: Condens.
  Matter} \textbf{2007}, \emph{19}, 255208\relax
\mciteBstWouldAddEndPuncttrue
\mciteSetBstMidEndSepPunct{\mcitedefaultmidpunct}
{\mcitedefaultendpunct}{\mcitedefaultseppunct}\relax
\EndOfBibitem
\bibitem[Nishimura et~al.({2008})Nishimura, Kobayashi, Ohoyama, Kanno, Yashima,
  and Yamada]{Nishimura2008}
Nishimura,~S.-i.; Kobayashi,~G.; Ohoyama,~K.; Kanno,~R.; Yashima,~M.;
  Yamada,~A. \emph{{Nature Mater.}} \textbf{{2008}}, \emph{{7}},
  {707--711}\relax
\mciteBstWouldAddEndPuncttrue
\mciteSetBstMidEndSepPunct{\mcitedefaultmidpunct}
{\mcitedefaultendpunct}{\mcitedefaultseppunct}\relax
\EndOfBibitem
\bibitem[Amin et~al.({2008})Amin, Maier, Balaya, Chen, and Lin]{Amin2008}
Amin,~R.; Maier,~J.; Balaya,~P.; Chen,~D.~P.; Lin,~C.~T. \emph{{Solid State
  Ionics}} \textbf{{2008}}, \emph{{179}}, {1683--1687}\relax
\mciteBstWouldAddEndPuncttrue
\mciteSetBstMidEndSepPunct{\mcitedefaultmidpunct}
{\mcitedefaultendpunct}{\mcitedefaultseppunct}\relax
\EndOfBibitem
\bibitem[Balluffi et~al.(2005)Balluffi, Allen, Carter, and
  Kemper]{balluffi2005kinetics}
Balluffi,~R.~W.; Allen,~S.~M.; Carter,~W.~C.; Kemper,~R.~A. \emph{{Kinetics of
  Materials}}; John Wiley \& Sons: New Jersey, 2005\relax
\mciteBstWouldAddEndPuncttrue
\mciteSetBstMidEndSepPunct{\mcitedefaultmidpunct}
{\mcitedefaultendpunct}{\mcitedefaultseppunct}\relax
\EndOfBibitem
\bibitem[Li et~al.({2008})Li, Yao, Martin, and Vaknin]{Li2008}
Li,~J.; Yao,~W.; Martin,~S.; Vaknin,~D. \emph{{Solid State Ionics}}
  \textbf{{2008}}, \emph{{179}}, {2016--2019}\relax
\mciteBstWouldAddEndPuncttrue
\mciteSetBstMidEndSepPunct{\mcitedefaultmidpunct}
{\mcitedefaultendpunct}{\mcitedefaultseppunct}\relax
\EndOfBibitem
\bibitem[Amin and Maier(2008)Amin, and Maier]{Amin20081831}
Amin,~R.; Maier,~J. \emph{Solid State Ionics} \textbf{2008}, \emph{178},
  1831--1836\relax
\mciteBstWouldAddEndPuncttrue
\mciteSetBstMidEndSepPunct{\mcitedefaultmidpunct}
{\mcitedefaultendpunct}{\mcitedefaultseppunct}\relax
\EndOfBibitem
\end{mcitethebibliography}

\providecommand*\mcitethebibliography{\thebibliography}
\csname @ifundefined\endcsname{endmcitethebibliography}
  {\let\endmcitethebibliography\endthebibliography}{}

\end{document}